\documentclass[10pt,onecolumn,a4paper]{IEEEtran}

\usepackage{epsfig}
\usepackage[cmex10]{amsmath}
\usepackage{amsfonts,amssymb}

\usepackage[latin1]{inputenc}

\usepackage{algorithm}
\usepackage{algorithmic}
\usepackage{graphicx}

\usepackage{color, verbatim}
\begin{document}

\title{Network Coding Meets Multimedia: a Review}

\author{Enrico Magli, Mea Wang, Pascal Frossard, Athina Markopoulou}

\maketitle

\begin{abstract}
While every network node only relays messages in a traditional communication system, the recent network coding (NC) paradigm proposes to implement simple in-network processing with packet combinations in the nodes. NC extends the concept of ``encoding''  a message beyond source coding (for compression) and channel coding (for protection against errors and losses). It has been shown to increase network throughput compared to traditional networks implementation, to reduce delay and to provide robustness to transmission errors and network dynamics. These features are so appealing for multimedia applications that they have spurred a large research effort towards the development of multimedia-specific NC techniques. This paper reviews the recent work in NC for multimedia applications and focuses on the techniques that fill the gap between NC theory and practical applications. It outlines the benefits of NC and presents the open challenges in this area. The paper initially focuses on multimedia-specific aspects of network coding, in particular delay, in-network error control, and media-specific error control. These aspects permit to handle varying network conditions as well as client heterogeneity, which are critical to the design and deployment of multimedia systems. After introducing these general concepts,
the paper reviews in detail two applications that lend themselves naturally to NC via the cooperation and broadcast models, namely peer-to-peer multimedia streaming and wireless networking.
\end{abstract}

\section{Introduction}
\label{sec:intro}

Network coding (NC) is one of the recent breakthroughs in communications research. It has first been proposed in \cite{Ahlswede:2000ht} and potentially impacts all areas of communications and networking. In a nutshell, the conventional networking paradigm is based on routing, where packets are forwarded towards the destination by each network node. NC is based on the idea that intermediate network nodes process packets in addition to relaying them. Processing typically involves generating and sending new output packets as linear combinations of the incoming ones. It can be done at different levels of the classical protocol stack (transport/application, MAC/IP, physical layer, and joint processing involving multiple layers). This new paradigm entails a radical change in the design of communication networks and applications.

Multimedia delivery techniques heavily rely on the transport and network layers of the communication protocol stack. Multimedia communications have been optimized to fully exploit the traditional protocols and maximize the quality of service. With the new NC paradigm  in networks with source, path and peer diversity, the traditional coding and communication methods have to be rethought in terms of the way packets are created, combined and delivered. In particular, multimedia applications could clearly benefit from the improvements offered by NC in terms of augmented throughput, reduced vulnerability to packet erasures and ease of  deployment in large scale distributed systems.

The NC concept has been initially devised in the information theory community. The gap between the information theoretic models and the models of real multimedia applications is fairly large. This has generated a lot of interest in developing NC schemes that yield significant benefits in multimedia applications, especially  in peer-to-peer (P2P) streaming and wireless communication systems. In particular, multimedia applications pose unique and important challenges depending on the application scenario, such as large amounts of data, stringent timing constraints,  heterogeneity in the packet importance as well as in the client set. Therefore, practical NC algorithms have to be specifically designed  to cope with these challenges under several additional constraints from realistic settings.

First, the decoding complexity has to stay reasonable for realtime decoding. Second, the algorithms need to be scalable and robust to handle the dynamics of the sources, the intermediate nodes and the paths in a large network infrastructure. Finally, the NC solutions must be adaptive to accommodate various timing constraints in delivering multimedia content, as these are among the most fundamental issues to be addressed in a realistic multimedia system. All these issues call for an appropriate design of the distributed communication protocol and effective coding strategies. Similarly to the traditional multimedia communication framework, the most promising solutions in NC systems certainly rely on end-to-end system optimization based on a joint design of the source and network code, where timing or complexity constraints are added to the data communication problem.

The objective of this paper is to review recent work in NC-based multimedia communication and to provide a comprehensive survey of the problems that have been fully or partially solved, along with the open research questions. The paper guides the readers through these problems in the following way. In Section \ref{sec:theory}, we briefly review the theory of NC, we summarize the main theoretical results, and we put them in the context of practical multimedia communications applications. In Section \ref{sec:coding}, we review the specific NC techniques that have been developed to tackle the issues mentioned above. Our analysis covers in-network error control, as well as media-specific error control. The former refers to NC techniques that have the objective of maximizing {\em network} performance from a ``multimedia'' point of view, {\em e.g.} minimizing delay. Media-specific error control introduces {\em distortion} into the NC setting, employing network-oriented source coders based on layered and multiple description coding. Low-complexity NC techniques are also considered. Sections \ref{sec:p2p} and \ref{sec:wireless} describe two specific application domains of NC principles, namely P2P and wireless networks. These are important because they  naturally lend themselves to NC due the cooperation and broadcast models. In particular, Section \ref{sec:p2p} is concerned with the use of NC for P2P media streaming. It describes in detail how NC can radically change the way P2P streaming systems are designed, leveraging a coded representation that facilitates content distribution and streaming. It further includes a discussion of NC for both live and on-demand streaming. Section \ref{sec:wireless} is dedicated to the applications of NC in wireless networks, considering various topologies, traffic scenarios and coding schemes. It first describes how NC can be used to improve throughput in general wireless single and multiple unicasts. Then, media-specific issues are considered, and augmented versions of general-purpose NC solutions are discussed, which explicitly take into account the media format and structure. NC-based collaboration among mobile devices is discussed, reviewing the relevant techniques and open problems. Finally, Section \ref{sec:conclusions} concludes the paper  and outlines open problems. It should be noted that the other papers \cite{overview,athina_eletter} also review multimedia applications of NC, and the survey in \cite{mehta} particularly focuses on wireless multimedia applications. In this paper we provide a more comprehensive overview, with particular emphasis on both the P2P and wireless streaming problems.

\section{Network Coding Framework}
\label{sec:theory}

\subsection{Basics}

NC is a new paradigm for sending information over networks. Instead of simply relaying packets (as it is the case with routing), network nodes can also combine incoming packets and send the resulting \textit{coded} packets to outgoing edges. While these operations do not generate new information in the network, they augment the diversity in the representation of the source information, with several potential benefits.

In particular, NC has been shown to improve on throughput, delay, resilience to packet loss, and system complexity, when properly applied in favorable scenarios. At the same time, these gains do not come for free: NC requires processing in the middle of the network, which may lead to additional delays, complexity and vulnerability to byzantine attacks \cite{anh_le}. Therefore, NC is not equally attractive in all application settings. For example, NC has not been implemented in classic wireline networks (because of the cost and complexity of introducing changes to high speed routers), while it has been successfully applied to peer-to-peer and wireless mesh systems, where it is feasible for nodes to process incoming packets. For this reason, NC is particularly well suited to multicast traffic scenarios (as is the case in peer-to-peer networks where all nodes cooperate to get the same file) and wireless networks (where the broadcast medium provides opportunities for combination of diverse representations of the source information). Furthermore, NC is theoretically well-understood in these two types of network systems.

Before discussing applications in more detail, we now spend time on describing the principles of network coding. The seminal paper that defined the beginning of the NC research is the work in \cite{Ahlswede:2000ht}. It considers a multicast session over a directed graph with lossless links and shows that, when operations at intermediate nodes are allowed, the maximum multicast rate is equal to the minimum min-cut from the source to each receiver. Essentially, if all receivers have the same min-cut from the source, network coding permits to achieve the min-cut capacity simultaneously for all the nodes. This capacity corresponds to the maximum flow rate that each receiver could get if it were alone in the network. Later, the work in \cite{Li:2003vx} showed that linear network coding operations are sufficient for achieving this maximum multicast rate, and the work in \cite{Koetter:2003kh} further fostered the area by confirming these results in an algebraic coding framework.

\begin{figure}[t!]
\begin{center}
\includegraphics[width=0.45\textwidth]{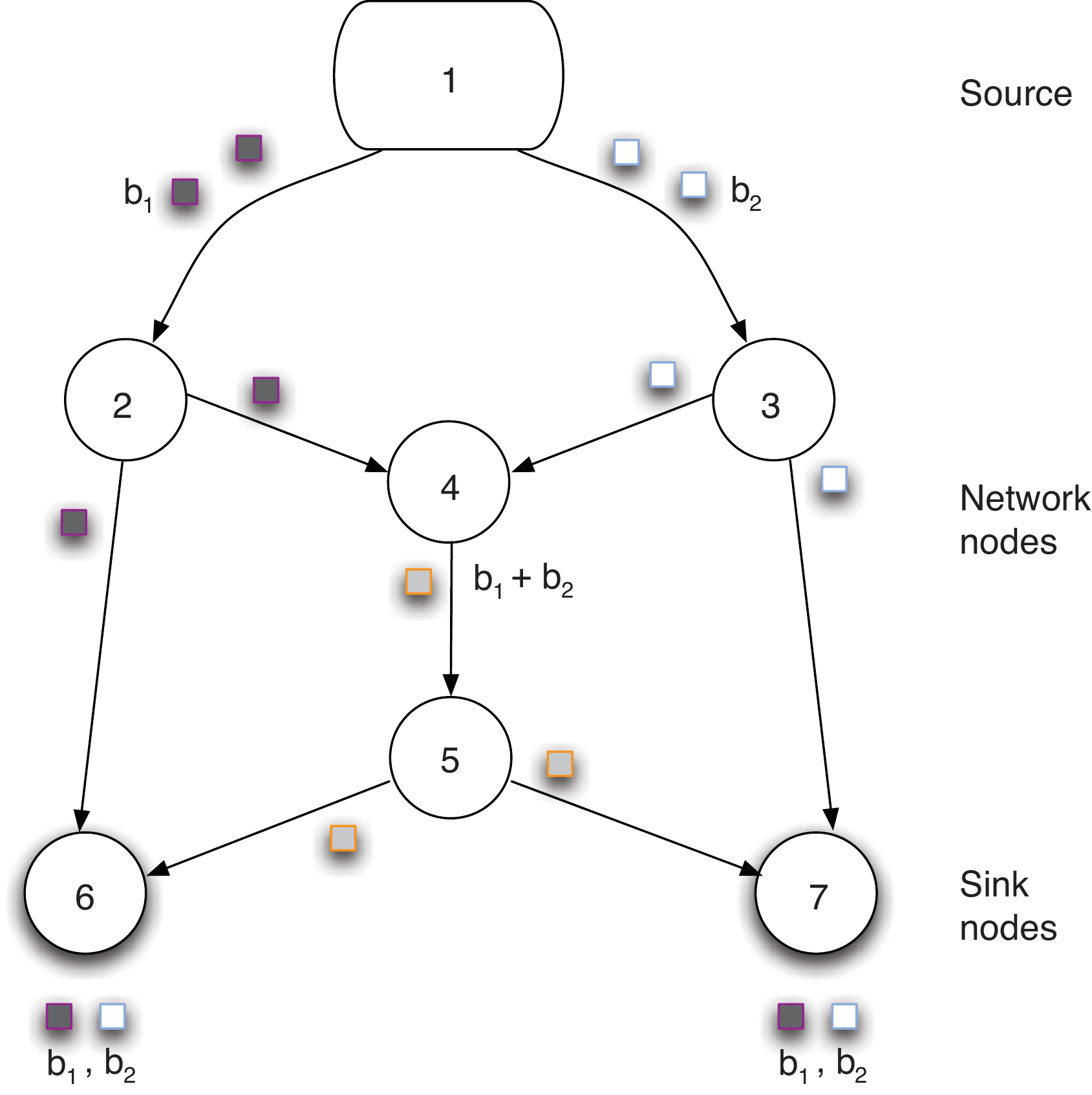}~\\
\end{center}
\caption{Butterfly network \cite{Ahlswede:2000ht} and one multicast session. There is one
source and two sinks. Links have a unit capacity, and zero delay.
{\em Intra-session} network coding is used on edge $4-5$.}
\label{fig:butterfly}
\end{figure}

The simplest example that demonstrates the key idea and the benefit of network coding in the multicast case is the butterfly network depicted in Fig. \ref{fig:butterfly}. The source $1$ sends two different symbols $b_1$ and $b_2$. The symbols are then transmitted towards the receivers\footnote{In this paper, the terms ``receiver'', ``sink'', or ``client'' are used interchangeably.}. Conventionally, two transmission slots are necessary to send both symbols through the bottleneck link between nodes $4$ and $5$. However, when the symbols are combined at node $4$, one transmission slot is sufficient to transmit the coded information that is useful to both sink nodes $6$ and $7$. Each of them receives the coded symbol (from the bottleneck links) and one of the original symbols (from the side links). The receivers can then decode both symbols $b_1$ and $b_2$. This simple example demonstrates that NC requires fewer time slots to convey the source information, compared to a routing solution. Network throughput is increased, as indeed each receiver achieves its min-cut, which is equal to 2. Moreover, the delivery delay is reduced by the combination of symbols at the intermediate node.

\begin{figure}[t]
\begin{center}
\includegraphics[width=0.45\textwidth]{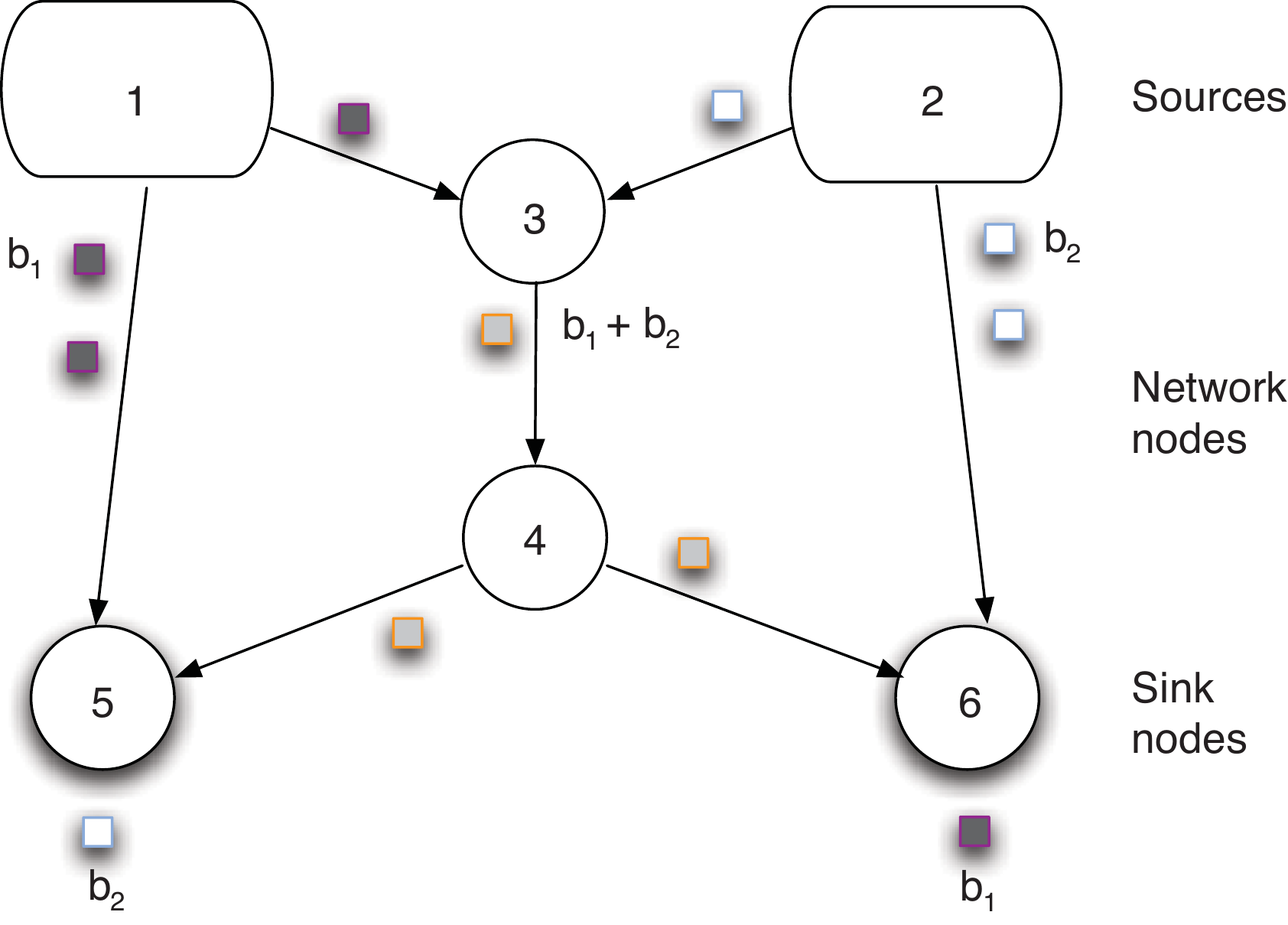}~\\
\end{center}
\caption{Modified butterfly network \cite{Ho:2008wg} and two unicast sessions ($1$-$6$ and $2$-$5$).
 All links have a unit capacity, and zero delay.
Inter-session network coding combines symbols from the two different sessions at node $3$.}
 \label{fig:mbutterfly}
\end{figure}

The main multicast theorem and the above example refer to {\em intra-session network coding}, because the symbols $b_1, b_2$ belong to the same multicast session\footnote{The term ``session'' refers to one source and one or more receivers interested in that source.}. Over the last decade, network coding has been applied to many network and traffic scenarios\footnote{A ``traffic scenario'' consists of one or more sessions taking place at the same time on the same network.}, beyond a single multicast session over lossless graphs. In particular, it has been extended to {\em inter-session network coding}, which refers to coding packets from different (unicast or multicast) sessions that share the network resources. Furthermore, network coding in networks with erasures and/or broadcast links have also been studied.

The simplest example that demonstrates inter-session coding is the modified butterfly example shown in Fig. \ref{fig:mbutterfly}. The sources $1$ and $2$ send information to receivers $5$ and $6$, respectively; the key difference with Fig.\ref{fig:butterfly} is that each receiver is only interested in receiving its intended symbol, not in both symbols. The combination of symbols from both sessions with \textit{Inter-Session network coding} at node $3$ efficiently uses the central bottleneck link, since the transmitted information is useful to both clients. The clients receive both coded symbols and the original symbols that are used together for decoding the information of interest in only one time slot. In contrast, routing needs two time slots to deliver the same information to the receivers in this network. In general, inter-session NC can be more efficient than a solution where the concurrent sessions are allocated different subgraphs, with intra-session NC in each subgraph \cite{Koetter:2003kh} \cite{Traskov:2006jx}. However, defining the optimal inter-session network coding operations is still an open problem, and suboptimal or heuristic approaches are used in practice.

\begin{figure}[t]
\begin{center}
\includegraphics[width=0.45\textwidth]{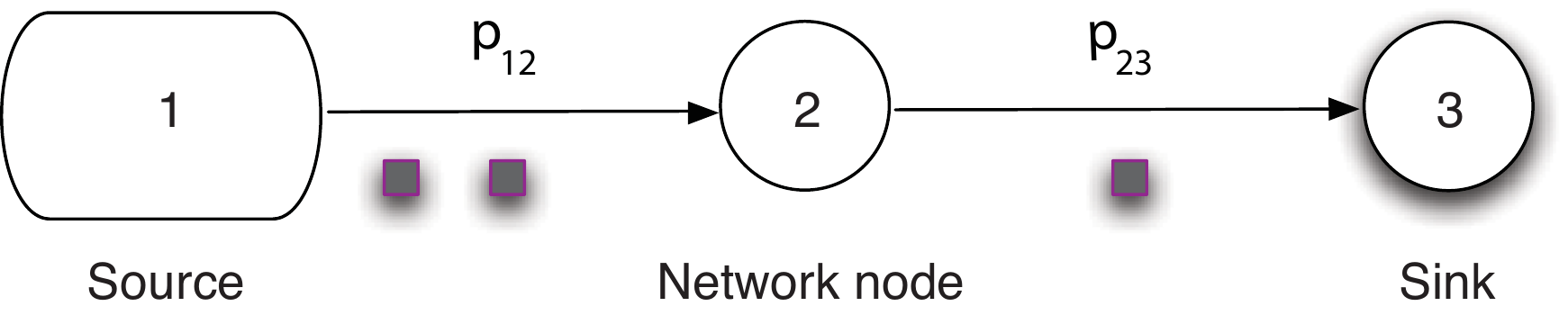}~\\
\end{center}
\caption{Two-link tandem network. Toy example of network with one source node and one sink node connected by lossy channels via a network node. Error resilience can be increased when NC is implemented in node $2$.} \label{fig:tandem}
\end{figure}

NC is also helpful for transmission over lossy channels, where it can help to increase the resilience to data loss. We illustrate this advantage by the simple example of a two-link tandem network given in Figure \ref{fig:tandem}. A source transmits information towards a sink node via two lossy links with symbol erasure probabilities $p_{12}$ and $p_{23}$. One could apply channel coding at the source, with a channel rate being computed based on the end-to-end loss probability. This leads to a communication rate of $(1-p_{12})(1-p_{23})$. It can be shown that, if the source implements a channel code for the first link, and then if node $2$ decodes and re-encodes with a channel code adapted to the reliability of the second link, the system can achieve a transmission rate of $\min (1-p_{12}, 1-p_{23})$; this is usually larger than $(1-p_{12})(1-p_{23})$ and hence better than an end-to-end channel coding strategy. But this form of in-network processing usually involves large delays due to the decoding operations in the network nodes. Interestingly enough, it is possible to achieve the same transmission rate of $\min (1-p_{12}, 1-p_{23})$ with NC \cite{Lun2006}. Node $2$ can simply combine packets as in intra-session NC in order to achieve capacity on the two-link tandem network, without decoding and recoding packets in the network nodes. It provides an effective and low-delay solution to combat losses with help of network nodes close to the failure points. It is interesting to note that this approach bears some resemblance with rateless or Fountain codes \cite{Shokrollahi06} \cite{Luby02}, where effective combinations of packets have led to error-resilient communications with a low overhead.

The above examples illustrate the benefits of NC in different communication problems that are commonly encountered in media streaming applications. The theoretical limits of NC are well studied in the field of network information theory. Interested readers may refer to \cite{ITNC,Yeung:2010wa,Ho:2008wg,Fragouli:2011wk,Fragouli:2006fc,Koetter:2003kh} for an in-depth understanding of NC foundations. In this paper, we concentrate on the application of NC for solving multimedia communication problems.

\subsection{Algebraic Network Coding}

We give now more details on the network coding principles, from the perspective of the algebraic NC framework. We consider general communication systems that are composed of source nodes, network nodes and sink nodes, which are connected by channels that are potentially lossy. We represent the system as a directed graph $\mathcal{G} = \{\mathcal{V},\mathcal{E}\}$, where the vertices $\mathcal{V}$ represent the different nodes in the network. The set of edges $\mathcal{E}$ consists of arcs between nodes and denotes the links in the network.\footnote{For simplicity, links here are considered point-to-point, but the framework can be extended to include wireless broadcast channels (using hyperarcs and a hypergraph).} The NC problem first estimates a set of rates to be used on each edge for data delivery. This leads to the definition of the coding subgraph, {\em i.e.}, the subset of the capacitated graph to be used for data delivery. The problem is then to determine the proper NC assignment among nodes in the coding subgraph.

The general framework of NC is not limited to specific types of network nodes. Most of the theoretical results, however, use linear codes, where the output symbols are linear combinations of the input symbols. Linear codes are sufficient to achieve capacity in the lossless multicast problem \cite{Li:2011gr,Li:2003vx}, but insufficient for optimal inter-session NC, where pathological counter-examples exist.
However, they lead to a computationally tractable yet effective solution; therefore, linear network coding is the only type of coding used in practice and is thus of interest to this paper.
Algebraic linear network coding can be summarized as follows \cite{Li:2011gr, Koetter:2003kh}. Consider an edge $e=(u,v) \in \mathcal{E}$ with $head(e)=u$ and $tail(e)=v$. Then, a symbol $y(e)$ that is sent over an edge $e$ emanating from node $u$ can be represented as a linear combination of the symbols $y(e')$ that have been received by the node $u$ over the set of edges $e' = \{e \in \mathcal{E} | tail(e)=u\}$. It can be expressed as

\begin{equation}
y(e) = \sum_{e'} \mathbb{\alpha_{e}}(e') y(e') ,
\end{equation}

where each edge is associated with a vector of local coding coefficients $\mathbb{\alpha_{e}}$ whose length is equal to the indegree of node $u$. The coded symbols are actually combinations of the source symbols $x_i$, with $i \in {1.. N}$. Indeed, it can be shown by induction that the successive encoding operations in the different nodes of the network result in the following relation

\begin{equation}
y(e) = \sum_{i}^N \mathbb{\beta_{e}}(i) x_i
\end{equation}

The vector of global coding coefficients $\mathbb{\beta_{e}}$ for the edge $e$ can be written as a function of the local encoding coefficients $\mathbb{\alpha_{e}}$ used at the different network nodes. It has a length equal to the number of source symbols, $N$. The sink nodes decode the linear system of equations generated based on the set of received packets, with classical techniques such as Gaussian elimination, in order to recover the source symbols.

While several works have been conducted towards optimal linear code design ({\em i.e.}, the proper selection of the coefficients $\alpha_i$) based on the network constraints \cite{Jaggi:2005bg}, the coding coefficients are typically chosen randomly based on Random Linear Network Coding (RLNC). Remarkably, it has been shown in \cite{random_linear_NC} that taking the coefficients at random from a uniform distribution over a Galois field of size $q$ leads to good performance. As $q$ becomes large, the probability that randomly drawn encoding vectors yield linearly dependent packets becomes arbitrarily small. Furthermore, since the random drawings can be performed independently at each network node, there is no need for centralized knowledge of the topology, nor of the buffer states: this leads to a fully distributed algorithm. RLNC is thus particularly interesting in distributed systems where it leads to easy deployment without the need for coordination between nodes. In practice, a common choice of the Galois filed size is to set $q=256$. It is also possible to take a smaller field size, {\em e.g.}, $q=2$ as in \cite{cope}. The advantage of this latter choice resides in reduced complexity, which is very appealing in a wireless scenario for example.

\subsection{Practical Network Coding}
\label{sec:practical_NC}

In practical implementations for multimedia communications over packet networks, the  NC operations described in the previous subsection are applied to full data packets and not on bytes or elementary symbols independently. In particular, a packet is treated as a vector of symbols (typically one byte each), and the same coding coefficients are applied to all symbols in the packet with the coding operations on packets being done symbol-wise.

The design of practical network coding systems also implies that the sink nodes have to receive sufficient information for decoding the NC packets. This is solved in \cite{Chou:2003um} by tagging each packet with the set of coefficients that describe the coding operations, so that the sink node can perform the appropriate decoding operations. In practice, a header that conveys information about the coding coefficients (typically the global encoding vector $\mathbb{\beta_{e}}$ ) is added to each packet. The global encoding vector entails a rate overhead, whose size depends on the number of source symbols and the size of the finite field.

\begin{figure}[h]
\centering
\includegraphics[width=8cm]{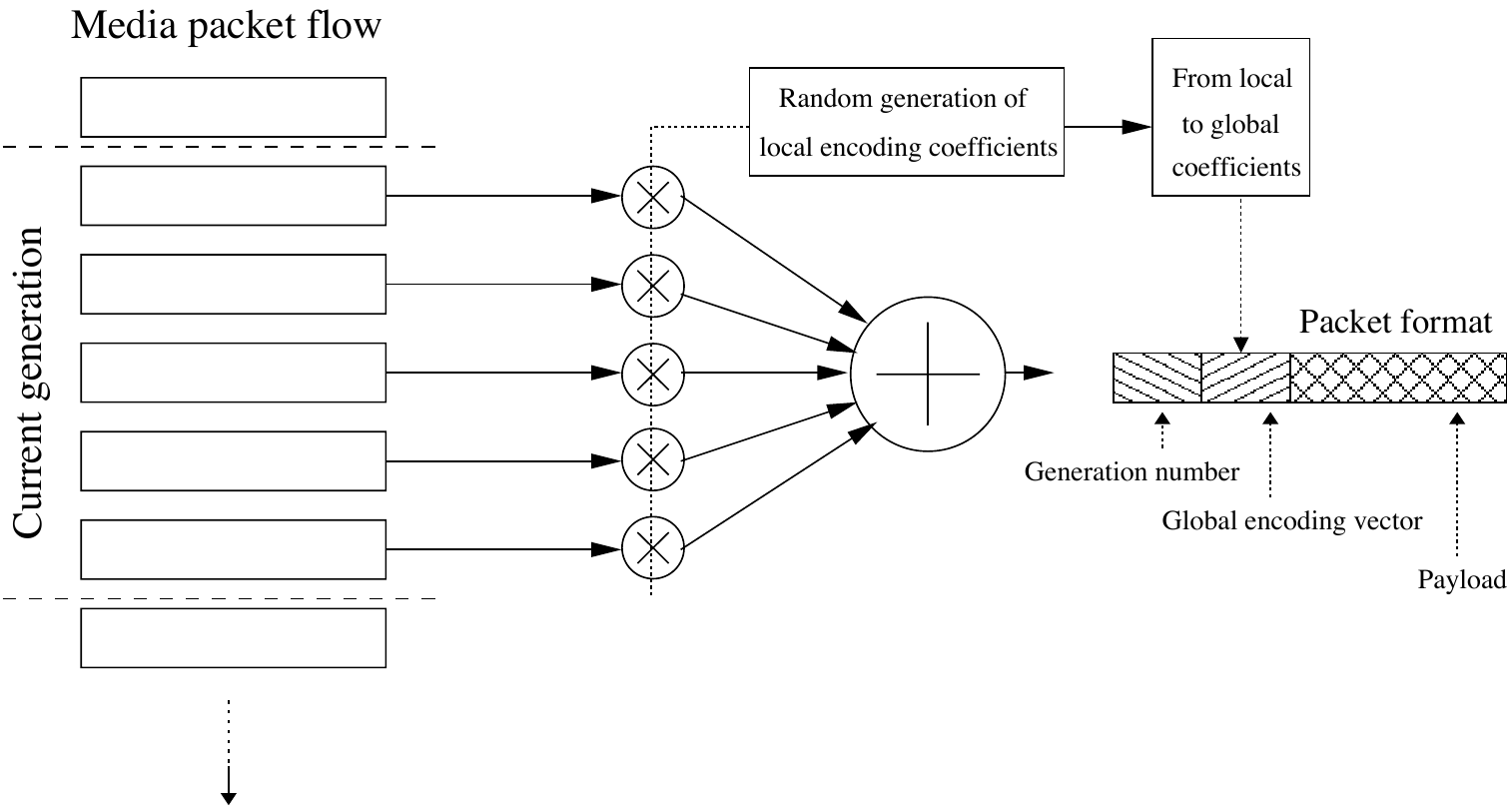}
\caption{Practical network coding system. Packets from a generation are encoded together, and the encoding coefficients are conveyed in the header of the resulting packet.}
\label{fig:practical_NC}
\end{figure}

An important difference between NC theory and practical settings lies in the timing constraints that are inherent in multimedia communications applications. Real networks transmit packets asynchronously and media streaming sources continuously transmit new packets that have to be decoded before their expiration deadlines. Therefore, packets belonging to a multimedia stream are generally grouped into generations of size $L$ ({\em i.e.}, sets of $L$ time consecutive multimedia packets) and network coding operations are applied to packets of the same generation \cite{Chou:2003um}. Each packet must also be tagged with the generation number, and each node may have to keep multiple buffers in order to separately process packets of different generations. The overall process is sketched in Figure \ref{fig:practical_NC}. We finally remark that the values of the generation size ($L$) and of the coding field $q$ play an important role in the performance of the system, leading to a trade-off between the header size, the computational complexity, and the decoding performance.

Equipped with the main principles for practical NC, we discuss in the next sections how NC solutions can be deployed in media streaming systems. We review solutions for data communication problems where data packets are heterogeneous in terms of their importance in  multimedia quality, where timing and computational complexity pose stringent constraints to the media delivery, or where the network topology changes during the session. We survey  different tools and algorithms that potentially benefit from the inclusion of NC in multimedia communications applications.

\section{Network coding adapted to multimedia}
\label{sec:coding}

Section \ref{sec:theory} has discussed the main aspects of NC for multimedia communications. There is, however, a gap between such models and multimedia applications in several aspects.

\begin{itemize}
\item In practical networks, packet losses may occur due to congestion, and large delays may render packets useless in multimedia applications. These aspects are not taken into account in the graph-based network model introduced in Section \ref{sec:theory}.
\item The computation of optimal NC schemes usually considers that the network topology or the nodes' buffer content is known {\em a priori}. However, in many multimedia applications such as peer-to-peer streaming, the topology is rarely known by any peer, and the buffer content changes over time. Moreover, the network is generally non-static and of large-scale.
\item The network model does not consider aspects such as delay jitter and bandwidth fluctuations, which are challenging issues for
multimedia streaming.
\item NC is all about increasing the available bit-rate, while in general multimedia applications are evaluated in terms of rate {\em and} distortion (or other suitable utility metrics).
\end{itemize}

The objective of this section is to review design techniques for NC, with a specific focus on the requirements posed by practical multimedia applications. We consider the practical NC framework defined in Section \ref{sec:practical_NC} and discuss methods for designing NC algorithms that are tailored to
multimedia streaming systems. In addition to improvement in throughput,
we pay a special attention to error resiliency and to algorithms that adapt NC operations to the video information in data packets. Finally, we review methods that address the delay and complexity constraints in multimedia applications.

\subsection{In-network error control}
\label{sec:feq_arq}

If a node does not receive sufficiently many packets due to congestions or delays, decoding of the content is not possible; this leads to dramatic quality degradations. Applying Forward Error
Correction (FEC) codes to the media amounts to adding redundancy symbols for
loss recovery. Since most multimedia communications occur in packet network
environments, losses typically occur in the form of packet erasures. Erasure
correction codes are a nice and viable solution, {\em e.g.}, Reed-Solomon codes,
low-density parity-check codes, turbo codes, as well as the more recent digital
fountain codes such as LT codes \cite{Luby02} or Raptor codes
\cite{Shokrollahi06}. The latter are based on the cascade application of a FEC
block code, followed by an LT code. This results in linear encoding and
decoding complexity that is attractive for NC.

These concepts can be extended to a distributed environment with multiple senders and
receivers, particularly when the intermediate nodes are allowed to apply NC. This leads to the concept of ``network error correction'', which has become increasingly popular in the information theory community \cite{nec_1,nec_2}. Practical coding solutions have mostly employed rank-metric codes \cite{silva_rank}. In these codes, codewords are matrices, as opposed to vectors, and the {\em rank} is employed as a distance metric between two codewords, {\em i.e.}, the rank of the matrix difference. Rank-metric codes have become popular for storage applications, where information is stored in two-dimensional arrays but errors are typically confined to one or few rows or columns. Gabidulin codes are the rank-metric analog of Reed-Solomon codes, and are probably the most studied rank-metric codes \cite{gabidulin,roth}. These codes are interesting for NC, as they can be used to solve a distributed erasure correction problem of the form $Y=AX+Z$, where the received packets $Y$ are modeled as combinations of the source packets $X$ using a linear transformation matrix $A$, plus a matrix $Z$ of error packets. It should be noted that, while there have been many recent theoretical advances in the area of network error correction, practical coding solutions have yet to be proposed.

In \cite{MingquanWuJSAP05}, the concept of  ``network-embedded forward error correction'' has been proposed. Conventional multicast systems apply FEC codes in an end-to-end fashion, that is, only the source nodes and the destination nodes are involved in the coding and decoding processes. These are delivered to the receivers, which attempt to perform FEC decoding in order to recover the original data. The idea behind network-embedded FEC is to have a few enhanced ``supernodes'' that perform FEC decoding and re-encoding, instead of simply relaying packets. These nodes act as signal regenerators: they are able to locally recover from information losses and thereby to significantly improve the overall performance, at the price of added complexity. In \cite{MingquanWuJSAP05}, a greedy algorithm is proposed to select the near-optimal locations of the enhanced nodes. The main conceptual difference between network-embedded FEC and NC lies in the decoding-recoding approach. In network-embedded FEC, every enhanced node has to wait until enough packets are received in order to be able to decode and then re-encode a data segment. Therefore, it is not possible for an enhanced node to {\em immediately} forward re-encoded packets on its outgoing links, which may induce delays in the system. While network-embedded FEC can be seen as a very
simple form of NC, it is not as good as a true NC solution, which would immediately re-encode and transmit packets without additional delay.

Channel coding techniques have also been employed in \cite{ThomosACM07,LTS-ARTICLE-2009-051} to perform NC on multimedia sessions in overlay networks. Rateless codes are used to construct an NC scheme in which the source packets are initially encoded by the server using a non-systematic Raptor code and then disseminated in the overlay. Each node schedules packet
transmissions according to the  amount of available bandwidth. If the number of received packets for a given media segment exceeds the available outgoing bandwidth, then only a part of the received packets are forwarded. Alternatively, if the available bandwidth is larger than the number of received packets, a node forwards the received packets, and then creates and forwards new linear combinations, leading to full exploitation of the available resources. In generating linear combinations of received symbols, one has to make sure that the new symbols are linearly independent to the original ones in order to avoid unnecessary redundancy. In \cite{LTS-ARTICLE-2009-051}, this has been taken care of by using combinations of only two symbols, and making sure that the same symbol is not used in multiple combinations. In general, this scheme is very effective computation-wise, but symbols cannot be recombined repeatedly, as this would generate linear dependencies in the new symbols. The scheme might thus present limitations for very large-scale networks.

Finally, another class of error control solutions is based on employing NC in the context of packet retransmission. In particular, it has been shown that existing transport protocols such as TCP can be augmented by NC, leading to significantly improved performance \cite{NC_TCP}. Similarly, NC has been employed in wireless networks in order to recover from packet losses. These aspects are discussed in more detail in Section \ref{sec:wireless}.

\subsection{Media-specific error control}
\label{sec:svc}\label{sec:mdc}

Media-specific error control tackles the problem of considering distortion, in addition to rate, in the NC design. This is achieved by employing media compression techniques that exhibit graceful quality degradation upon occurrence of packet losses.

Similarly to joint source and channel coding or in general joint optimization among multiple layers of the communication protocol stack in client/server scenarios, information about content availability can be used in network error control algorithms in order to improve resource utilization. The video content can be arranged in classes of different importance in terms of video quality, or in multiple versions that gracefully increase the video quality. Such arrangements provided by techniques like scalable or multiple description video coding permit to design effective content-specific error control solutions.

\begin{figure}[h]
\centering
\includegraphics[width=8cm]{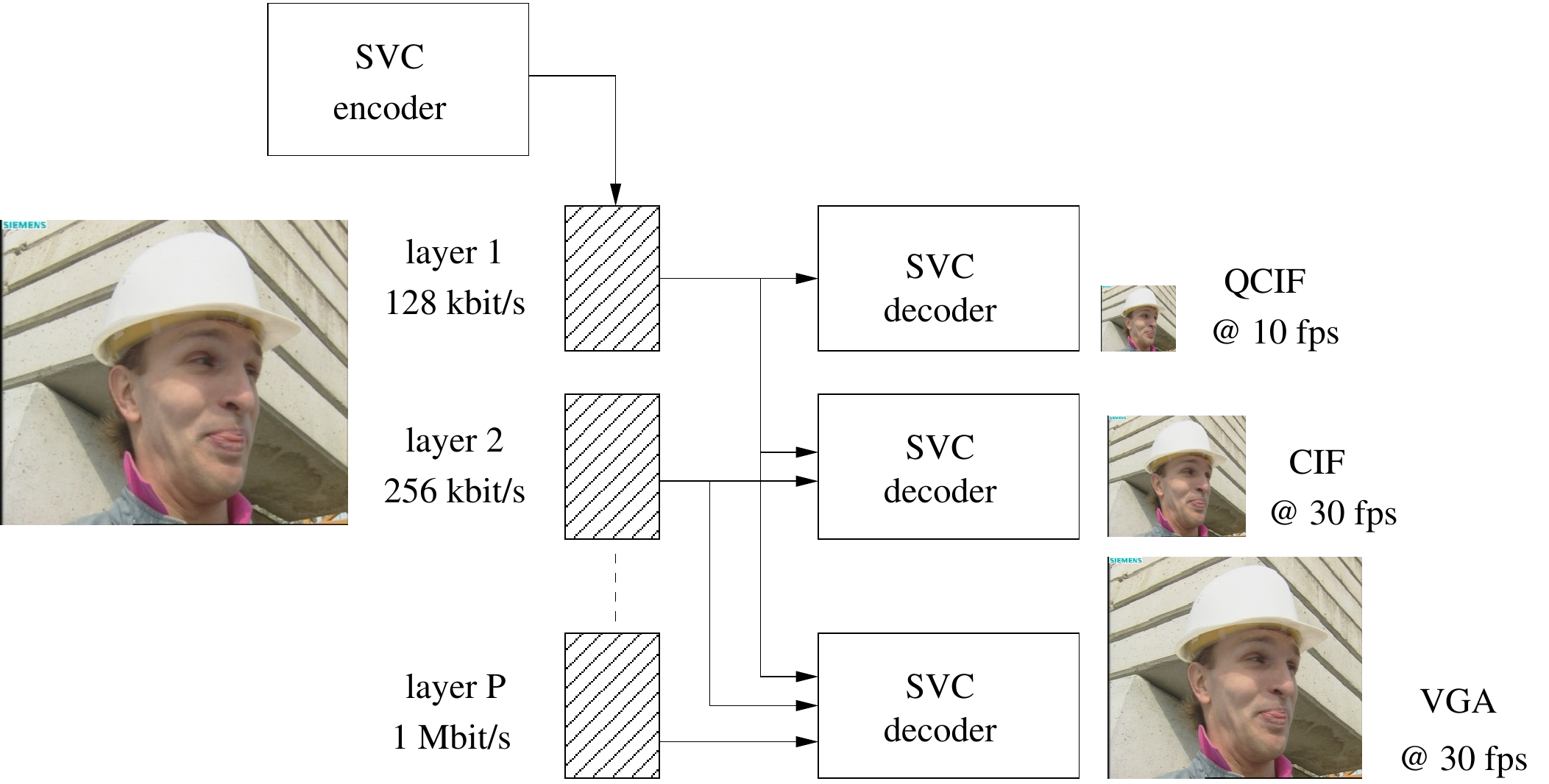}
\caption{Scalable video coding with $P$ layers at increasing
bit-rates. Decoding is done at different cumulative rates, yielding a video at
increasing frame-rate and spatial resolution.} \label{fig:svc}
\end{figure}

In scalable video coding (SVC) \cite{svc_overview}, multiple ``quality layers'' are created in order to adapt to the different capabilities of heterogeneous clients. The quality layers are typically decoded in a hierarchical way, as shown in Figure \ref{fig:svc}. In a $P$-layer system, the first layer provides a baseline level in terms of video quality and/or spatial and temporal resolutions. Additional enhancement layers can be used to improve on quality, resolution, or a combination thereof. SVC should be coupled with unequal error protection techniques in order to provide a more reliable delivery of the most important layers. This can be obtained in various ways, {\em e.g.}, allocating a lower FEC code rate or more retransmission opportunities to the first layers. A very flexible scheme for the transmission of scalable coded media is represented by priority encoding transmission \cite{albanese}. In this scheme, one can specify priorities for different media segments ({\em e.g.}, quality layers), and packets are created as a combination of source and erasure correction code symbols, such that the most important layers can be recovered from few packets, with more and more enhancement layers being recovered as more packets are received.

Interestingly, the unequal error protection principle can be easily extended to NC. In particular, in \cite{Chou:2003um} it is proposed to implement priority encoding transmission in a distributed way, exploiting NC packet recombinations to generate the FEC symbols in a distributed way. Thus, relatively few linearly independent packets are sufficient to recover the base layer during the NC decoding process, while more packets are needed to recover the enhancement layers. A similar approach has been proposed in \cite{nguyen_cheung}. Given $P$ quality layers, $P$ different ``types'' of packets can be created, where a packet of type $i$ contains linear combinations of source packets from layers 1 to $i$, but not packets from layers $i+1$ to $P$. In this way, packets of type $i$ can be used to decode any layer from 1 to $i$. The result is similar to \cite{Chou:2003um}, in that the base layer can be recovered from a limited number of packets. The number of packets of each type can be modulated in order to reach a desired probability of successful decoding for a given layer.

The approach proposed in \cite{stankovic_mmsp2010} is also similar in spirit to the above unequal error protection schemes. Multimedia sources encode data with scalable coding principles. Hence, the first part of the compressed file is more important, as it contains the base layer. Then a NC scheme is designed in such a way as to achieve higher probability of successful decoding for the first part of a media segment. This is done by dividing the media into layers and creating NC packets that are linear combinations of packets belonging to the first $i$ out of $P$ layers. The authors in \cite{stankovic_mmsp2010} investigate the problem of designing distributions of packet types that can lead to optimal quality. They provide an analytical evaluation of the decoding probability for each layer, enabling accurate control of quality of experience through the design of a suitable distribution. In \cite{nc_mgm}, the multi-generation mixing approach has been proposed. With this technique, coding can be performed over different generations, which facilitates decoding in case of incomplete generations. Decoding of a generation can be performed independently if sufficiently many packets have been received, and jointly with other generations otherwise. Note that multiple generations with inter-generation decoding as in \cite{nc_mgm} can also be interpreted as representing the media as one single generation divided into layers, with inter-layer combinations and joint decoding as in \cite{stankovic_mmsp2010}.

The problem of delivering scalable content has also been addressed in \cite{walsh_weber}. In scalable coding with NC, the receiver has to wait until the complete segment has been received before starting the decoding process. The authors of \cite{walsh_weber} have designed a delay-mitigating code that can be applied to a layered source in such a way that the first and most important layers are decoded earlier and  forwarded to the application layer without any delay. This is achieved through the concatenation of a short-length delay-mitigating code and a network code. In general, the ability to receive and decode the layers in order comes at the expense of a larger overall overhead.

Finally, the unequal error protection problem has been addressed from a packet scheduling perspective in \cite{prioritized}. Unlike previous approaches that are based on the design of codes yielding unequal protection, the authors in \cite{prioritized} have defined coding decisions that are adapted to the capabilities of different receivers. The receiving peers request packets from different layers and prioritized transmission is achieved by varying the number of packets from each layer that are employed in the NC combinations. The authors propose a distributed optimization algorithm that minimizes the expected distortion in the overlay network.

As an alternative to hierarchical video layers, multiple description coding (MDC) \cite{puri} can also generate multiple video streams, where the ``descriptions'' are independently decodable and hence do not require any unequal error protection. The combination of NC and MDC has been given comparatively less attention than SVC has been. In \cite{woods_vcip2009}, the authors have proposed to use multiple description wavelet-based video coding based on FEC codes in the framework of priority encoding transmission \cite{puri}. The multiple description scheme
employs random codes as in \cite{Chou:2003um}. A similar approach has been developed in \cite{wang_kuo}. In this case, the objective is to optimize the redundancy of each layer in an interleaved packet-based priority encoding transmission scheme. The authors develop redundancy allocation algorithms and finally apply their scheme to H.264/SVC.

\subsection{Delay and complexity constraints}
\label{sec:constraints}

We describe now NC techniques that specifically address the delay and complexity constraints in  multimedia applications. In general, NC can reduce delay as the reduced number of transmission opportunities employed to deliver the content often translates into a reduced delay experienced by the receiver. At the same time, NC is a block-coding operation where a block represents a generation; hence the encoder has to wait for a given number of symbols before these symbols can be processed and transmitted. This introduces a ``minimum'' delay, which may even be an unbounded quantity in the  case where rateless codes are used, although the delay can be bounded with very high probability. Since NC is essentially a distributed FEC code, the trade-off between performance, delay and complexity is somewhat similar to that for regular non-distributed block coding. High-performance coding for short block lengths (and hence low delay) is possible, but it generally requires to employ ``dense'' linear combinations, {\em i.e.}, combinations such that most of the coefficients $\alpha_i$ are
nonzero. This leads to a code whose encoding and decoding complexity is large, and may be undesirable for applications in wireless networks and with battery-powered computing devices.

Low-complexity codes can be designed based on random coding theory, {\em e.g.}, low-density parity-check codes \cite{Gallager} or digital fountain codes \cite{Luby02,Shokrollahi06}. However, these codes typically require large block lengths to operate close to capacity, which leads to large delay. It is possible to employ a random code over a short block, but generally this  yields a less efficient code. As a consequence, the receiver will have to wait for a possibly large number of packets before decoding, as some of the received packets are likely to be linearly dependent. Another influential factor is the size of the Galois field that the code is constructed on. In general, small field sizes lead to lower complexity codes, and are amenable to precomputation of multiplications so that the encoding  only consists of table look-up operations. However, it is known that the smaller the field size, the larger the probability of receiving linearly dependent combinations; hence, a larger bandwidth overhead is needed to perform successful decoding. In short, there are several  trade-offs to determine for achieving bandwidth efficiency, complexity and delay.

A few specific solutions have been proposed to lower the complexity of NC while still achieving good performance. In \cite{fiandrotti_mmsp2011}, two modifications have been proposed. First,
the linear combinations are picked directly from the partial decoding results generated during the Gaussian elimination process. Second, the degree of the linear combinations is kept low by combining only a limited number of packets at a time. In \cite{fiandrotti_icme2012}, band codes have been proposed, which permit fine tuning of the complexity of NC schemes with very little performance penalty. A similar problem has been addressed in \cite{thomos_degree}. The authors propose to employ modified degree distributions at the NC encoders, calculated in such a way that a desired distribution is obtained at the decoders for low-complexity decoding. Finally, the complexity problem can also be addressed at a higher level. In \cite{cleju}, it has been proposed that only a limited number of appropriately-selected nodes of an overlay network perform linear combinations while the other nodes simply relay their incoming packets. This minimizes the transmission delay incurred by NC.

Recent implementations of network coding libraries, such as NCUtils
\cite{ncutils}, Tenor \cite{Tenor} and KoDo \cite{kodo} have demonstrated that it is possible to achieve low delay and high encoding/decoding rates on commercial platforms, including resource-constrained smartphones. For example, the work in \cite{microcast-mobisys12} implements a cooperative video streaming system on Android phones, using the NCUtils library \cite{ncutils}. Experiments demonstrate that, for a field size of 8 and generation size of 25, encoding and decoding rates are in the order of 5Mbps and 25Mbps; this is sufficient for video streaming purposes in practice. This was achieved through a combination of optimizations such as progressive encoding/decoding, selection of the generation size and field size, table-lookups for multiplications and XOR for additions, random selection of only a subset of the coefficients, implementation in Java and native C++ on Android phones.

In summary, there are different ways to leverage NC and improve media streaming performance, as summarized in Tab. \ref{tab:mm}. A first class of  techniques are designed to improve networking by properly designing linear combinations and packet scheduling policies in order to optimize decoding performance. Another class of NC techniques exploit specific media features such as scalable and multiple description coding in order to maximize the quality of  experience. Overall, several papers have demonstrated their usefulness in ad-hoc scenarios, but in-network techniques still lack a unified and comprehensive framework. Moreover, although there is a good understanding of the basic principles of NC for multimedia streaming, optimal solutions have yet to be found. For example, NC for SVC and multiple description coding is very challenging, as it involves two aspects at the same time, {\em i.e.}, designing an NC scheme that requires a small overhead and designing suitable packet scheduling policies for the different substreams. None of these problems has been fully solved, and there is still a lot of room for research in these areas. As NC is making its move towards mobile battery-powered devices, the topic of low-complexity NC schemes is also very timely and relevant.

We provide in the next sections more details about specific  algorithms in two of the most popular frameworks for multimedia NC, namely peer-to-peer systems and wireless scenarios.

\begin{table*}[tbph]
\caption{NC techniques for multimedia applications}
\label{tab:mm}
\begin{center}
\begin{tabular}{|p{1.6in}|c|p{2.5in}|l|}
\hline
{\bf Classes of Proposal} &  {\bf Section} & {\bf Key Benefits}  & {\bf References}  \\
 \hline
In-network error control & \ref{sec:feq_arq} & Increased goodput, reduced delay, increased error resilience &  \cite{MingquanWuJSAP05,ThomosACM07,LTS-ARTICLE-2009-051}\\
\hline
Scalable video coding & \ref{sec:svc} & Improved quality of experience under packet losses (requires unequal error protection), handling of heterogeneous clients (bit-rate/quality, display resolution, temporal resolution) & \cite{Chou:2003um,nguyen_cheung,stankovic_mmsp2010,nc_mgm,walsh_weber,prioritized} \\
\hline
Multiple description coding & \ref{sec:mdc} & Improved quality of experience under packet losses (does not require unequal error protection) & \cite{woods_vcip2009,wang_kuo}\\
\hline
Low-complexity network coding & \ref{sec:constraints} & Improved battery life, reduced computational delay, ability to handle computation-heterogeneous clients & \cite{fiandrotti_mmsp2011,fiandrotti_icme2012,thomos_degree} \\
\hline
\end{tabular}
\end{center}
\end{table*}

\section{Peer-to-peer Multimedia Applications}
\label{sec:p2p}

The requirements of P2P live multimedia streaming applications have marked a
significant departure from P2P file sharing. In {\em P2P streaming}, like PPLive
\cite{pplive}, a peer plays the video while receiving the content from the
source, which requires consistent and sustainable throughput over time in order
to maintain a smooth playback. The streaming content is made available by one or
more servers, known as the {\em streaming sources}, and is partitioned into {\em
groups of frames (GOFs)} representing $t$ seconds of the playback. The smallest
unit of data that is being transmitted from the server to a client is commonly
referred to as a {\em segment}, which encompasses $n > 0$ GOFs. As participating
peers, end hosts contribute their upload bandwidth capacities to serve other
peers. Consequently, the load on dedicated streaming servers is significantly
reduced. Thus, each peer can potentially be a server for the content it has
received. When peer A receives a segment from peer B, peer A is peer B's {\em
downstream} peer, and peer B is the {\em seed} of this segment for peer A. The
successful deployments of PPLive \cite{pplive} and CoolStreaming \cite{ZLLY05}
have demonstrated the feasibility of P2P multimedia streaming over the Internet.
The salient advantage of P2P streaming is the {\em scalability} of a streaming
session since the upload capacities on streaming sources are no longer the
bottleneck.

In the above context, NC has the advantages of imposing less bandwidth demand on
servers and offering better playback quality. It also leads to shorter initial
delay as well as more robustness to network dynamics \cite{largescale}. The
equal importance of all NC packets naturally balances the bandwidth demand on
each peer according to its upload capacity. Furthermore, NC transmits small
blocks of segments rather than segments in their entirety, which leads to better
utilization of peer upload capacity and smaller waste of bandwidth when discarding
blocks of obsolete segments. The effective bandwidth utilization naturally
leads to very short initial buffering delay, especially during the flash-crowd
scenario. In addition, it significantly reduces the bandwidth demand on the streaming
server. Ultimately, the bandwidth saving enables a better scaling of the network. The
feasibility of NC is based on a special protocol design in which the random
coding coefficients are either embedded in each of the transmitted coded blocks
or can be reproduced using the random generator seed used in the segment. In this section, we
first review the P2P infrastructure for multimedia
streaming since NC operates on top of a P2P streaming protocol. We then discuss
the applications of NC in P2P live streaming, followed by the applications of NC
in P2P Video-on-Demand (VoD). We conclude this section with a discussion of new
research directions in applying NC for multimedia streaming.

\subsection{P2P approaches for multimedia streaming}

Stemming from the success of P2P file sharing, a few state-of-the-art P2P
streaming algorithms have been proposed to alleviate bandwidth demands on the
streaming sources. There are mainly two structures in P2P streaming systems:
push-based and pull-based. The push-based solutions, also known as the tree-based
approaches \cite{splitstream, msr:mutualcast, coopnet, resilient}, are based on
the philosophy of IP multicast. In such a paradigm, peers are organized into one
or more multicast trees rooted at the source. The source ``pushes'' {\em
segments} to all descendants among the trees. Although the push-based approaches
lead to short delays in distributing the content, it is not generally employed
in real-world streaming applications, mainly due to the complexity and
difficulty in maintaining the structured network topology in the
presence of peer dynamics.

In sharp contrast, the pull-based approaches, also known as the mesh-based approaches, do
not enforce any rigid structure among the peers \cite{bullet, ZLLY05, ZXZY05,
chunkyspread, chunkyspread1, prime}. Instead, connections are established
dynamically based on the content availability at each peer. The content
exchanged in this approach is best described as a ``data-driven'' or
``swarming'' style of multicast. In data swarming, each peer advertises to its
neighbours the  availability of segments in its buffer, and the neighbours explicitly
request segments as needed. The primary advantages here are simplicity in
maintaining peer connectivities and robustness in dynamic networks. Nonetheless,
the delay in delivering the live content to each participating peer is
inevitably increased due to the periodical exchange of segment
availability among peers.

Regardless of the actual algorithm and protocol in use, the P2P infrastructure
poses unique challenges in designing a scalable and high performance multimedia
streaming systems. First, peers may arrive at and depart from a session in
unpredictable ways.  Second,
peers must communicate with each other in order to avoid receiving redundant
content. This is commonly referred to as {\em data reconciliation}
and often results in excessive communication overhead and a lack of adaptivity
to network dynamics. Third, the heterogeneity in peer upload capacities makes it
challenging to {\em balance the load} among peers without a central
administration. Last, but not least, the diverse network delays and jitters
require peers to buffer a sufficient number of segments before the start of the playback in
order to maintain a smooth rendering. The time required for this buffering is
commonly referred to as {\em initial delay} or {\em playback delay}. In the next
two sections, we discuss how NC can help to address these challenges.

\subsection{Network Coding for P2P Live Streaming}
\label{sec:p2plive}

NC has been one of the key enabling technologies that marks a significant design shift in
P2P multimedia streaming. The highly visible {\em Avalanche} project \cite{GR05} has demonstrated the first practical use of NC in
P2P file sharing systems. It showed that  all pieces of information are
treated equally with NC. Without the need to identify and distribute the ``rarest
piece'' first, the control protocols are greatly simplified since the
need for data reconciliation is eliminated. The simplified protocol along with the coded data
swarming concept lead to a faster download rate with a minimal coding operation
overhead. It has been shown that the performance benefits provided by NC
correspond to a 2-3 times higher throughput compared to a system that does not use NC at all
\cite{GR05}. In addition, the residual bandwidth at peers can be utilized to
transmit small encoded blocks.

Due to the great performance gain brought by NC in P2P file sharing systems,
network researchers and developers are extending its application to P2P live
multimedia streaming systems. The benefits and trade-offs of NC in P2P live
streaming have been studied in \cite{WL07,lava}. These studies shows that the use of
NC leads to better bandwidth utilization and better resiliency to network
dynamics. Furthermore, under limited bandwidth, NC-enabled streaming systems
offer better performance than conventional streaming systems do.

\begin{figure}[htp]
\vspace{-3mm}
\begin{center}
 \includegraphics[width=0.3\textwidth]{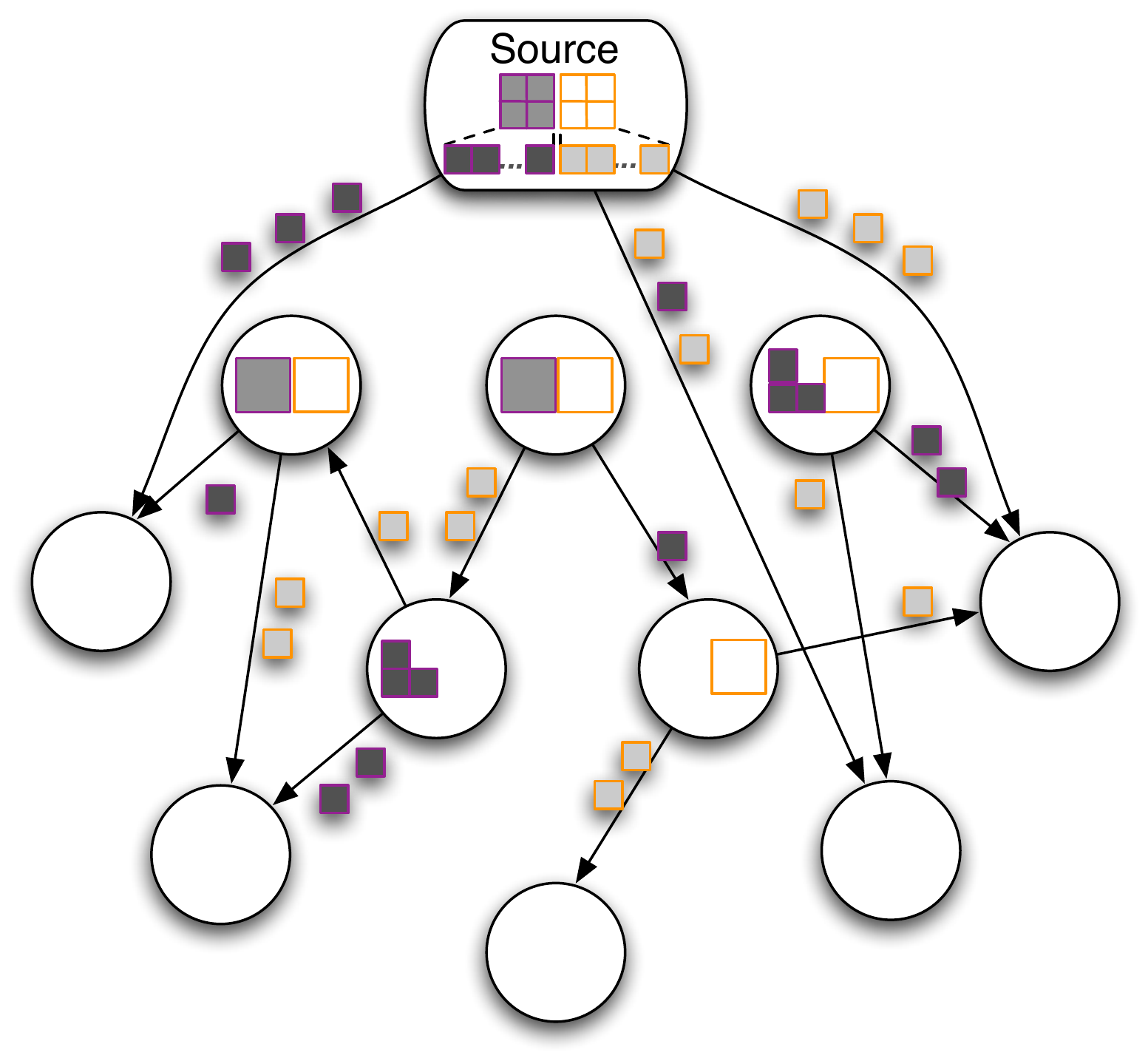}
 \caption{Network coding in P2P streaming systems.  Each segment is divided into small blocks that are coded using NC.  Small coded blocks are being exchanged among peers.}
 \label{fig:P2P-example}
\end{center}
\vspace{-4mm}
\end{figure}

We analyze now in more details the benefits of NC in P2P live streaming. With NC, video segments are
divided into blocks and then encoded, as shown in Figure~\ref{fig:P2P-example}. Peers exchange small coded blocks rather
than the large raw segments. Hence, a peer may download coded blocks of a
segment from multiple seeds simultaneously without using any protocol to
coordinate the seeds. Once a peer completely decodes the segment, it informs all
its seeds by sending short messages to terminate the transmission of the
segment. Without using NC, a peer has to explicitly request individual segments
from the seeds.

\begin{figure}[htp]
\begin{center}
 \includegraphics[width=0.5\textwidth]{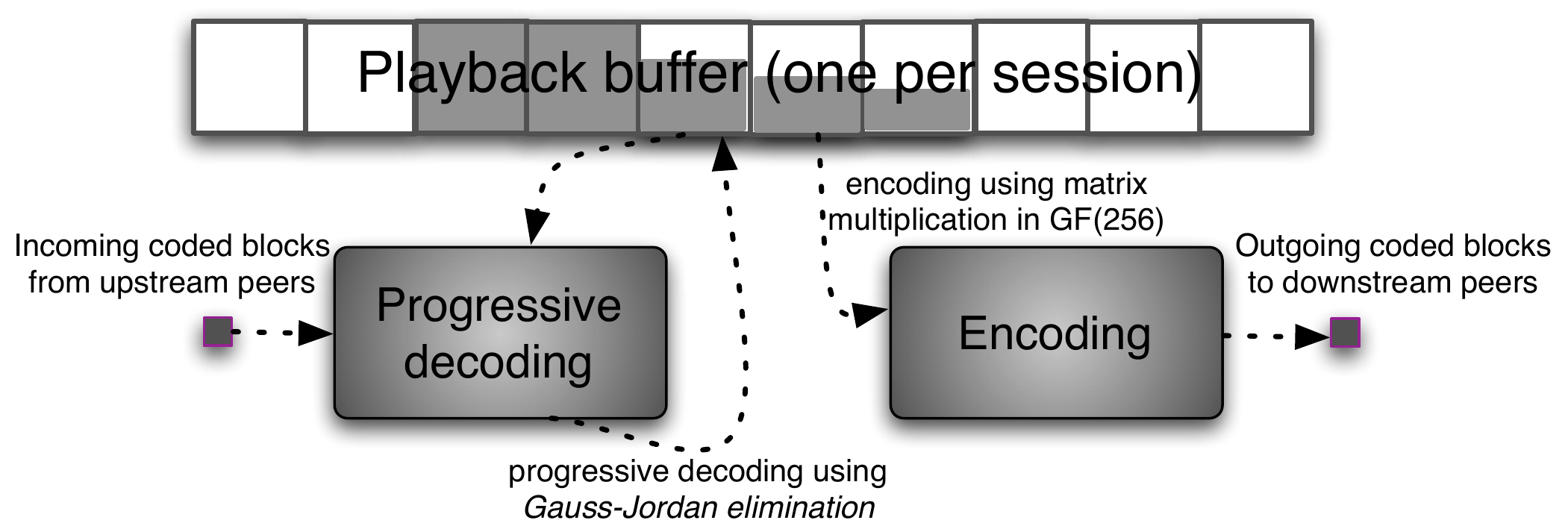}
 \caption{Progressive decoding for P2P streaming}
 \label{fig:P2P-buffer}
\end{center}
\vspace{-4mm}
\end{figure}

As mentioned in Section~\ref{sec:coding}, the delay introduced by NC operations can increase the delivery time of video segments, which
is critical in multimedia streaming sessions. In order to minimize this
overhead, progressive decoding has been introduced in \cite{WL07,lava}. With
progressive decoding, as illustrated in Figure~\ref{fig:P2P-buffer}, a peer can
start decoding as soon as the first coded block is received. Then, it {\em
progressively} decodes the new coded blocks as soon as they are received.
In this process, the decoding time overlaps with the time required to receive
the coded blocks, thus mitigating the coding delay within the transmission time of
a segment.

The application of NC fundamentally changes the way segments are shared among
peers, and the uniform importance of each coded block within a segment
greatly simplifies the data reconciliation process. A peer no longer relies on a
single seed for a given segment. It can simultaneously receive different blocks of a segment from
multiple seeds. Consequently, the system is more resilient to network dynamics
since the departure of one seed only affects a small portion of a segment, which
can be quickly recovered with data from other seeds. The delivery of small coded blocks also makes good use
of upload capacity, even for peers with very small residual upload
capacity. This effectively leads to faster delivery of segments and better
playback quality.

These fundamental changes encourage a re-design of the P2P live streaming system
to take full advantage of NC. To this end, the authors in \cite{r2}
have redesigned the P2P streaming protocol, $R^2$, which incorporates random NC with a
randomized push algorithm. Unlike the ``static'' push-based design using
predetermined trees in \cite{ZXZY05}, a peer in $R^2$ randomly chooses a
downstream peer and pushes coded blocks for a segment that is found missing without the need for any explicit request. We finally note that the $R^2$ design has been adapted by
UUSee \cite{UUSee10}, hence it represents the first real-world deployment of NC.

\subsection{Network coding for P2P Video-on-Demand}
\label{sec:p2pVoD}

{\em Video-on-Demand (VoD)} is a special case of live streaming, where the
constraint of synchronized playback among peers is relaxed. VoD users enjoy the
flexibility of watching arbitrary video segments at any time, including DVD-like
playback operations: play, pause, chapter selection, fast forward, and rewind.
P2P VoD has been widely deployed \cite{UUSee10, pplive, kangaroo, iplayer}, but it
has received less attention in the research community compared to P2P live
streaming services. Many works have focused on the ability to support
asynchronous playback but not the full set of VCR/DVD-like features. RedCarpet
\cite{redcarpet} is the first work incorporating NC into a P2P VoD system that
achieves a better playback quality with a simple protocol. Annapureddy {\em et
al.} \cite{qualityVoD} have then addressed the challenge of providing VoD using P2P
mesh-based network. They have also shown that it is possible to provide high quality
VoD using an algorithm combining NC, optimized resource allocation and overlay
topology management. SonicVoD \cite{SonicVoD09} and UUSee \cite{UUSee10} are the
two latest implementations of NC in real P2P streaming applications.
Measurements from UUSee \cite{UUSee10} confirm that NC can significantly
reduce the initial delay and server bandwidth cost in the streaming
architecture. They also show that, when multiple
seeds proactively push coded blocks to neighbouring peers, more blocks than
necessary for decoding may be transmitted, resulting in a waste of
resources. This calls for an early breaking mechanism, proposed in
\cite{UUSee10}, so that a peer can effectively stop the seeds from sending coded
blocks when it is about to receive sufficiently many blocks for decoding.

Despite the success of NC, the performance gain provided by NC has yet to be
quantified. SonicVoD \cite{SonicVoD09} has provided a comparison study to this end,
but the evaluation results are questionable. NC enabled VoD
applications cannot achieve smooth playback until the bandwidth provision is equvalent to  at
least twice the streaming rate, and the baseline VoD applications can never
achieve smooth playback. In reality, the streaming protocol cannot be that
inefficient in its bandwidth utilization. Therefore, the comparison results are
yet to be confirmed. More in-depth work on NC for P2P VoD is to be performed.

\subsection{What's Next?}

Although the application of NC in P2P multimedia streaming has received a lot
of attention, there are still several issues to be investigated. Such
research is certainly encouraged by the growth of P2P traffic. This growth raises
many concerns for ISPs and content providers, and it opens new research
directions towards system scalability and ISP traffic localization. In this section, we
briefly review ongoing work and future research directions in NC for P2P multimedia
streaming.

\subsubsection{Multimedia-based optimization}
\label{sec:optimizing}

An important research direction deals with the heterogeneity of the peers in P2P
systems, and the adaptation of video quality to the capabilities of each client.
It therefore becomes important to optimize NC for better video quality and
better network resource utilization. In order to accommodate the heterogeneity
in peer upload capacity, in \cite{multirate} the P2P streaming systems have
been modelled as a multi-rate multicast problem, and the optimal linear NC
assignment is computed using linear programming optimization. This solution
leads to the maximum aggregate rate among all participating peers in a streaming
session.

One of the distinct characteristics of multimedia content is that it can be
received in different forms and with different qualities depending on the
bandwidth availability. For example, scalable coding and multiple description
coding organize the content into multiple layers that can be decoded in a
flexible way, as has been shown in the previous section. The quality of the playback then typically depends on the number
of layers that have been received by individual peers. The feasibility of
combining NC with SVC has been explored in \cite{chameleon, shabnamLive}. In
\cite{chameleon}, the proposed system, {\em Chameleon}, groups the network
abstraction layer (NAL) units of SVC that have the same quality into a single packet. With NC, each packet is divided into smaller
blocks, just like the way segments are divided in both P2P live streaming
and VoD systems. Instead of transmitting raw packets, coded blocks of each
packet are shared among peers. Such a design extends the benefits that have been identified in
live streaming to the SVC-based streaming services. Chameleon shows up to
$2.5\%$ of improvement in playback skips and better resilience to peer dynamics.
In \cite{shabnamLive}, the proposed system utilizes both NC and SVC to better
support heterogeneous bandwidth among peers and to allow quick adaptation to
network churns.

Recently, a P2P multimedia streaming system has been proposed that
utilizes prioritized, randomized NC \cite{prioritized}. The original video
content is divided into classes of video frames and NC is applied differently in
each class. An optimal rate is then determined for each class on each network
link. For practicality, a greedy algorithm is proposed for computing the rate
allocation independently at each node. Using the video scalability property,
hierarchical NC \cite{nguyen_cheung} has been proposed to ensure that the base
layer of a layered bitstream is received with high probability. In addition,
a network coding algorithm has been proposed to solve the problem
for quality requirements of the receivers in multi-layered video streaming
\cite{ncMultiLayered}. In summary, optimization can further be performed at all
levels of the system, and the joint design of NC along with effective
rate allocation in the peers can lead to optimized delivery in heterogeneous P2P
environments.

\subsubsection{Scalability}
\label{sec:scalability}

The second important open issue in P2P multimedia streaming is scalability, that is, the
ability of the system to serve many users. It has recently been predicted that
the number of online video viewers will grow to more than 940 millions by 2013.
This clearly requires important scalability properties, which calls for a
redesign of video codecs, P2P protocols, hardware provisioning, and P2P traffic
management. In general, cooperation at all levels, from peers to ISPs to content
providers, is necessary in order to build a scalable multimedia streaming
system.

Based on existing deployments and research results, it is clear that P2P is a
promising approach for improving the scalability of multimedia streaming
systems. Furthermore, since P2P traffic constitutes $92\%$ of the cross-ISP
traffic, improving the multimedia streaming applications will not only
save resources to ISPs, but it will also reduce the redundant traffic across ISPs. To
this end, the work in \cite{friendly} examines the inter-ISP traffic incurred by P2P
streaming applications. It points out that fully localizing P2P streaming
traffic significantly degrades the overall streaming quality due to the lack of
diverse content among the peers. The solution to effective P2P streaming relies
on identifying the maximum level of traffic localization and on the design
of an ISP-friendly P2P streaming approach.

Interestingly enough, NC can help in localizing P2P traffic too. A
locality-aware network coding (LANC) P2P streaming system has been proposed in
\cite{LANC}. It uses locality information in both the topology construction and
the segment scheduling. LANC shows that the random property of
randomized NC increases the chance of finding useful data in nearby peers, which
in turn reduces the inter-ISP traffic. The simulation results show that LANC
can effectively reduce the inter-ISP traffic by $50\%$.

Finally, unlike cable and satellite services, multimedia streaming services over the
Internet is best-effort. The current Internet-based streaming are still not
comparable to the services offered by cable and satellite networks, especially
for HD TV and channel surfing experience. The potential of NC has
still to be unveiled. For example, the authors in \cite{layeredP2Plive}
have proposed a layered network coding solution for building collaboration incentives
among peers and improving content availability in P2P live streaming systems.
They utilize layered coding as a tool to generate incentives and combine it with
a specific network coding scheme to achieve efficient scheduling and delay
minimization. Such solutions might permit to get the best of the best-effort services, and to improve the quality of P2P streaming applications.

In summary, regardless of the internal design of the multimedia streaming
service, NC clearly improves bandwidth utilization and reduces the overhead in
the communication protocol.  These benefits consequently lead to
smoother playback and higher video fidelity. The work on NC applications in P2P multimedia
systems are recapped in Table \ref{tab:P2P}. Nonetheless, there remain many open
problems, unaddressed issues, and unexplored potentials in incorporating NC into
diverse systems for effective P2P multimedia streaming.

\begin{table*}[tbph]
\caption{NC Applications in P2P Multimedia Systems}
\label{tab:P2P}
\begin{center}
\begin{tabular}{|p{1.6in}|c|p{3.2in}|l|}
\hline
{\bf Classes of Proposal} &  {\bf Section} & {\bf Key Benefits}  & {\bf References}  \\
 \hline
NC for live streaming protocol & \ref{sec:p2plive} & Simplified protocol design, smoother playback, efficient bandwidth utilizes, shorter initial delay, and resilient to dynamics & \cite{WL07, lava, r2, UUSee10} \\
\hline
NC for VoD protocol & \ref{sec:p2pVoD} & Same as in live streaming, plus more responsive to seek operations & \cite{redcarpet, qualityVoD, SonicVoD09, UUSee10} \\
\hline
Combining NC and SVC & \ref{sec:optimizing} & Smoother playback, resilient to dynamics, and better support for heterogeneous systems & \cite{chameleon, shabnamLive}\\
\hline
Combining NC and layered coding & \ref{sec:optimizing} & Better adaptability and diverse viewing experience & \cite{prioritized, nguyen_cheung, ncMultiLayered} \\
\hline
New uses & \ref{sec:optimizing} & Maximized system throughput & \cite{multirate} \\
 & \ref{sec:scalability} & ISP traffic localization & \cite{friendly, LANC} \\
 & & Incentive building and content availability boosting & \cite{layeredP2Plive} \\
\hline
\end{tabular}
\end{center}
\end{table*}

\section{Wireless Multimedia Applications}
\label{sec:wireless}

Wireless environments lend themselves naturally to network coding,  thanks to the inherent overhearing and superposition capabilities of the wireless medium. The former (overhearing) provides side information to each next-hop node, which can then be used for decoding coded packets.  The latter (superposition) can be viewed as physical-layer network coding \cite{pnc1, pnc2, pnc3}, where coding is not introduced by  nodes but rather by the wireless medium itself \cite{wif-main, wif-coding}. In this section, we focus on overhearing, which is better understood today and already used for multimedia applications.

\begin{table*}[t!]
\caption{Representative Wireless Network Coding Scenarios \label{table:wireless-taxonomy}}
\begin{center}
\begin{tabular}{| c | c |c | p{1.2in}| p{1in}| p{1in}|}
\hline
 {\bf Traffic Scenario} &  {\bf Section} & {\bf Wireless Topology} & {\bf Coding Scheme} & {\bf Example Application}  & {\bf References}  \\
 \hline
   Single Unicast Session & V.A & Multihop &  source [+other nodes], intra-session & TCP over wireless & \cite{NC_TCP, medard-videocentric, medard-ita11}\\ 
   Multiple Unicast Sessions & V.B & multihop & in the middle, inter-session ({\em e.g.} COPE)  & Wireless mesh networks & \cite{cope, hs_JSAC09, nguyen_ncmedia, idnc-unicast}\\
   Single Multicast Session & V.C & one-hop downlink  & at  BS or AP, intra-session & Video Broadcast & \cite{ff1, ff2, Hua_P2P, aalborg1}\\ 
   Multicast Session(s) & V.D & device-to-device links & all nodes, intra [and inter-] session & Cooperative Mobile Devices & \cite{microcast-mobisys12, microplay-mobigames12, PictureViewer, GestureFlow}\\
  \hline
\end{tabular}
\end{center}
\end{table*}

We review wireless coded systems that have been proposed for  different traffic scenarios,
 network topologies and coding schemes, as summarized in Table \ref{table:wireless-taxonomy}.
Although  this taxonomy can be applied to both data and media traffic, we focus here on issues specific to media traffic. Given the large literature on wireless network coding, this review does not attempt to be exhaustive but tries only to highlight some representative schemes.

\subsection{Single Unicast}

First, let us consider a single unicast flow with intra-session coding.  Intra-session coding can be implemented at the end points and potentially also at selected nodes ({\em e.g.}, access points or mobile devices) within the network. Coding at the source requires neither changes in the network nor cooperations with other flows. Random network coding is the simplest form of network coding to implement at intermediate nodes.

In \cite{NC_TCP}, the authors have considered a single TCP flow transmitted over a single path with one or more lossy links.
The lossy links are typically wireless links but they can potentially be wired links as well. The authors have proposed a thin coding layer under TCP. At the source, this layer sends linear combinations of packets within a TCP-like window; the size of this window is adjusted based on feedback from the receiver. At the receiving end, this layer collects, decodes and delivers the packets to the TCP receiver; it also sends acknowledgements back to the source. The difference with TCP is that, instead of acknowledging individual packet sequence numbers, the receiver reports the dimensions of the subspace spanned by the vectors received so far, and the sender accordingly decides how many coded packets to retransmit. The goal of this network coding layer is to make the flow resilient to packet loss in a way that is transparent to TCP, thus avoiding retransmissions.
The key contribution of \cite{NC_TCP} is the incorporation of network coding  with minimal changes to the protocol stack.

Interestingly, these ideas are applicable not only to data but also to media flows. For example, the method in \cite{NC_TCP}  is directly applicable to HTTP streaming, which represents the majority of video streaming traffic today.
In \cite{medard-videocentric},  a strategy for network coding in wireless video transmission has been presented, which uses a
variety of coding approaches and adds feedback and device discovery to tailor the coding to the receivers.

The idea has also been applied to wireless video streaming over multiple paths.
In \cite{medard-ita11}, several connections to the same phone have been considered, {\em e.g.}, WiFi (which is considered  free but unreliable) and LTE (which is considered expensive but reliable), and coding has been performed across  these heterogeneous networks.
 The authors propose to use the more expensive network only when there is a risk of interruption in viewing video. This way, the scheme incurs the cost of the cheaper network, while at the same time achieving the performance of the best network. Approaches based on combining heterogeneous networks without NC suffer from the difficulty of having to coordinate across separate flows, which may require retransmission of specific packets across one or more networks.
 Coding approaches offer the flexibility  of combining data across different networks.

We would like to note that applying coding only at the end-points is a very restricted version of network coding, essentially similar to classic channel coding. However, some of these ideas gracefully generalize to allowing coding at intermediate nodes. For example,  consider a single flow going over a multi-hop network with lossy links (an example, for a tandem network of three nodes, has been discussed in detail in Section II and illustrated in Fig.\ref{fig:tandem}).
Allowing random network coding at intermediate nodes  is known to achieve the optimal rate with minimum delay \cite{loss1, loss2}. This is in contrast to alternatives,  such as end-to-end FEC (which adjusts the redundancy to the end-to-end loss rate, and thus cannot achieve the optimal rate), ARQ schemes (which achieve the optimal rate but with increased delay), or even schemes where intermediate nodes can decode and re-encode (which also achieve the optimal rate at the cost of additional delay).

\subsection{Multiple Unicasts}

Second, let us consider multiple unicast flows over multihop wireless networks, such as  wireless mesh networks. One could still use per-flow intra-session network coding as described before, but higher  benefit can be achieved using NC at all intermediate nodes and/or with inter-session coding.

Given that optimal inter-session network coding is an open problem, constructive approaches have been proposed  \cite{chou-unicast}, \cite{cope}, \cite{tiling}, \cite{Traskov:2006jx}, \cite{BFLY}.
One of the first and most influential practical wireless
network coding systems is COPE \cite{cope} - a coding layer between
the IP and MAC layers that performs one-hop, opportunistic
network coding for multihop wireless networks.
In COPE, each node combines packets from incoming unicast
flows using simple (bitwise XOR) coding operations.
Each node then broadcasts the coded packet and the next hop nodes
are able to decode using overheard packets from previous transmissions.
 This way, COPE effectively forwards multiple packets in a single transmission to improve throughput.
The scheme has been implemented in a WiFi testbed consisting of laptops  with madWiFi drivers; it has
been shown to significantly improve throughput  in a range of scenarios, especially for UDP traffic \cite{cope}.

COPE was designed to support any type of unicast flows. Next, we describe modifications and extensions of COPE proposed specifically for media flows.
The core design question in this setting is the {\em coding scheme}: which packets  to
select and code together from all possible subsets of packets at an intermediate node's queue?
In COPE,  a node always transmits the head-of-line packet and then greedily considers  each subsequent packet,
so as to ensure that the largest number of receivers can decode.  This policy increases
the information per packet transmission and thus the throughput.
However, when the transmitted flows are media streams, this FIFO greedy policy is
not necessarily the best choice. In particular, when coding packets together, one should take into account not only  the decodability  by as many neighbors as possible, but also the contribution of the packets to video quality.
In other words, when  media streams are transmitted, it is not only the number but
also the content of the packets that should
be taken into account by the network coding scheme.

The unequal importance of media packets and  the need for prioritization in transport mechanisms are well-understood in the media streaming community. Incorporating this idea into network coding schemes was first described for intra-session network coding  \cite{Chou:2003um} along the lines of priority encoding transmission \cite{albanese}, as described in Section \ref{sec:svc}.
This idea of prioritization among media packets was applied specifically in the context of COPE \cite{hs_JSAC09}. The algorithms build on top of COPE and  combine  several packets from different streams,  thus  increasing throughput. However, they also select which video packets to code by taking into account their importance in terms of video distortion and playout deadlines. Simulations for various topologies and traffic patterns have shown that the combined approach greatly improves video quality for the same  throughput.
Three algorithms (NCV, NCVD and NC-RaDiO) have been proposed in \cite{hs_JSAC09}, which employ different coding subsets of packets of different size, in order of increasing complexity and performance.
NCV (``Network Coding for Video'') adopts the greedy selection of COPE but considers a metric that captures distortion as opposed to the number of packets. NCVD is similar but is not greedy:  it considers all possible subsets of packets in the queue, which is a NP-hard problem. Finally, NC-RaDiO achieves the optimal performance within the rate-distortion optimized (RaDiO) packet scheduling framework \cite{radio}. Realistic simulations have shown that NCVD is able to achieve near-optimal performance with reasonable computational complexity for the small queue sizes that are typical of real-time media traffic.

Independently, in \cite{nguyen_ncmedia},
the RaDiO  framework \cite{radio} has also been applied
 to the transmission of multicast and unicast traffic  over a single broadcast wireless link when XOR network coding is allowed at the access point. The authors use a Markov Decision Process framework to optimize the policy at the access point,  {\em i.e.}, to decide whether to send a new packet, retransmit a lost packet, or transmit an XOR-ed packet.

In \cite{idnc-unicast},  inter-session network coding across two unicast sessions over an unreliable wireless channel has been studied. Each unicast session transmits a stored video file, whose packets have hard sequential deadline constraints. The authors characterize the capacity region for the transmission rates of the two unicast sessions under heterogeneous channel conditions and heterogeneous deadline constraints. Then, they develop ``immediately-decodable network coding'' (IDNC) schemes for controlling packet transmissions for the unicast sessions. They show the asymptotic optimality of the proposed IDNC scheme within the deadline constraints, when the file sizes are sufficiently large. In general, establishing the optimality of inter-session  network coding is non-trivial. An additional difficulty in the analysis of  \cite{idnc-unicast} stems from the heterogeneity of channel conditions and deadline constraints.

A scheme that combines {\em intra-session and inter-session network coding} for unicast flows over multi-hop wireless networks, I$^2$NC, has been proposed in \cite{i2nc}.  I$^2$NC builds on and extends COPE: it allows intermediate nodes to perform not only one-hop opportunistic inter-session  network coding as in COPE but also to add per flow redundancy using intra-session network coding. The combined approach has two benefits compared to COPE.  First, I$^2$NC is resilient to high loss rates, which are typical in wireless links. Second, I$^2$NC is sequence-agnostic  in the sense that intermediate nodes can operate using information only about the loss rates on the downstream links, as opposed to detailed reports about which packets have been received (as implemented in COPE). The scheme is also applicable to media flows and can be further optimized to take into account the loss resilience of media flows.

There is also a significant body of work on {\em cross-layer optimization} of rate-control, routing, scheduling and network coding, typically formulated within the network utility maximization framework.
 This general framework is applicable to multimedia flows as well if utilities and constraints are appropriately defined  to capture multimedia-specific considerations such as distortion and delay.
Next, we describe a few representative examples, which have focused specifically on video and on the interaction between video rate control and inter-session network coding.

The rate of multimedia flows typically varies over time, depending on the content and on the rate control. Bursty traffic patterns are well-known to not fully exploit the underlying coding opportunities because  packets from different flows must be present at the same node and at the same time in order for them to be coded together. This was originally noticed in the context of TCP but arises also in inter-session coding of media streams with time-varying rates. Two approaches have been proposed to deal with this problem.

The first approach \cite{delay-TCP, hs_icme09} proposes to delay packets at intermediate nodes in order to create more network coding opportunities; it faces an inherent tradeoff between delay and throughput. Intermediate nodes have control over how long to wait before coding or forwarding packets from different flows. At  one extreme if they forward packets immediately, they minimize delay but may also miss coding opportunities, thus  reducing throughput. At the other extreme, they can delay packets from one flow until packets from another flow arrive and code them together. The relation between delay and throughput has been analyzed based on random arrivals \cite{delay-TCP, hs_icme09, Sprintson1}, and assumptions on game \cite{Sprintson2} or adversarial \cite{Sprintson3} settings.

The second approach \cite{hs_pv09} proposes a cross-layer optimization of rate control at the higher  layer and network coding at the underlying layer in order to maximize the total utility (e.g., the overall expected video quality in the wireless streaming session). The authors of \cite{hs_pv09} exploit the observation that utility varies over time depending on the video content. As a result, two or more video streams can have different
rates over a short-time scale even though they may have similar rates over longer time scales. Additional delay is introduced in some video scenes and rate allocation
is optimized over longer time intervals in order to increase the NC opportunities and eventually maximize the  total utility.

\subsection{One-Hop Broadcast}

\begin{figure}[t!]
\centering
\includegraphics[scale=0.45,]{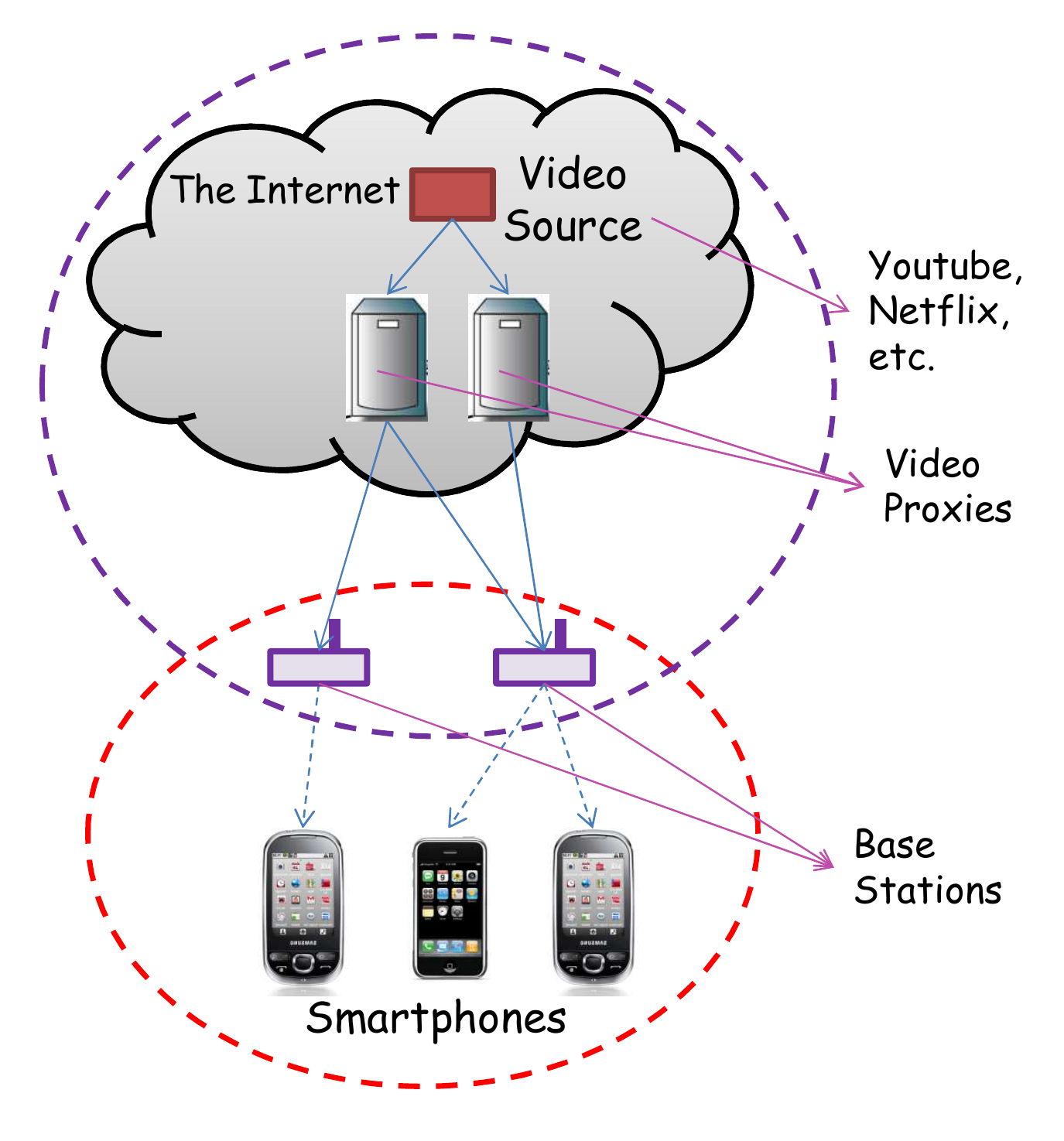}
\caption{{\em Overview of cooperative video streaming to mobile devices.} A group of smartphone users, within proximity of each other, are interested in viewing the same video at the same time. Each smartphone has Internet connection via cellular or WiFi connection.  When a user is interested in viewing a video, it connects to the video source ({\em e.g.,} YouTube or Netflix) via its base station, which may be the same or different for different users, depending on the provider their use. The proxy, in our system, is responsible for selecting the video rate and performing network coding .  Each smartphone  can receive packets from the source as well as from other smartphones in the neighborhood, through device-to-device (Bluetooth or WiFi)  links. Network coding can be used on the downlink (by the source or proxy) and/or on the local links (by the phones). }
\label{fig:smartphones}
\end{figure}

Many media applications are based on broadcast transmission, which perfectly matches the broadcast nature of the wireless medium. There is a significant body of work on wireless video broadcast where a single node broadcasts to several receivers.   (The broadcasting node can be thought of as the access point, base station, or video proxy, as depicted in Figure  \ref{fig:smartphones}). Coding over a broadcast link makes each transmission useful to multiple receivers at the same time.

The work in \cite{idnc-broadcast, idnc-broadcast-journal}  considers real-time streaming of stored video
using broadcast over the downlink of a single cell. The capacity of the system, subject to hard deadline
constraints, is derived for both uncoded and coded wireless
broadcast schemes. In \cite{ idnc-broadcast-journal},  it is shown that network coding is asymptotically
throughput-optimal, subject to the deadline constraints, when there are no more than three users  and  the file size is sufficiently large.
 Furthermore, numerical results show that IDNC schemes strictly outperform non-coding schemes not only in the asymptotic regime of large files but also for small files. One advantage of the immediately decodable network coding (IDNC) schemes is that they do not suffer from the decoding delay that is inherent to generation-based network coding protocols.

In \cite{ff1,ff2}, the goal is to achieve synchronized multimedia streaming to multiple iPhone devices. Broadcast transmissions  and re-transmissions from the base station are used.  In \cite{aalborg1} and \cite{Hua_P2P},  broadcast has been used on the downlink and ad-hoc links  have been used for error recovery.  In \cite{aalborg1}, a group of smartphone users are connected to the Internet via LTE links and packets are broadcast from the base stations using  the Multimedia Broadcast Multicast Services (MBMS). We note, however, that  although MSMS is standardized within  LTE, it is not currently implemented. In \cite{dejan_icme2012} it is proposed to integrate NC in the LTE/LTE-A radio access network protocol stack; the authors propose a MAC layer solution alternative to hybrid ARQ, which is particularly well suited to multimedia applications.

We further note that coding over a broadcast link with erasures is a well studied problem for which many coding schemes,  such as Fountain codes, have been previously proposed. The similarities of such schemes with linear network coding  can potentially be exploited and lead to combined or novel solutions for one-hop wireless broadcast.
For example,  a new code called BATS (``Batched Sparse'') has recently been proposed in \cite{bats-isit11}. BATS  enables a matrix generalization of fountain-like codes to be combined efficiently
with random linear NC.   A prototype has been built with a notebook computer broadcasting network-coded packets to iPod/iPad mobile
devices. In ongoing work, the authors investigate  the performance of BATS in different systems, including P2P and
wireless systems. Finally, we recall that, as discussed in Section \ref{sec:constraints}, coding for broadcast may increase the delay due to the block-coding or rateless coding operations; the delay may become unbounded, which is clearly undesirable for multimedia applications.

\subsection{Cooperative Mobile Devices}

Smart mobile devices, such as smartphones, tablets, and portable media players,  are widely used today. They are equipped with ever increasing computational, storage and networking capabilities. For example, smartphones today have several network connections (cellular, WiFi and most recently WiFiDirect and BlueTooth), which are  used to connect to
each other and to the Internet.
 Of interest here is media traffic to these devices, whether the content is generated/stored externally (on the cloud) or locally
(on one of the devices).

One example application is streaming video from YouTube or NetFlix to mobile devices. Although  the demand for such streaming services is exponentially increasing \cite{cisco-stats}, the downlink cellular connection  typically faces problems such as low or time varying rate, loss, and occasional outages. When a group of smartphones are  within proximity of each other,  they can potentially cooperate to deal with these problems, by establishing  device-to-device connections via WiFi or BlueTooth.
Cooperation between mobile devices is also motivated by the increasingly social aspect of mobile-to-mobile communication; mobile social networks are emerging as a field on its own \cite{distanceMatters, bubblerap, Boldrini2008, Yoneki2007, Jaho2009, Mashhadi2009}  and media content is routinely exchanged over these networks.

Cooperative mobile devices combine two orthogonal characteristics, namely cooperation and wireless transmission. First, as  discussed in Section \ref{sec:p2p},  network coding offers benefits in cooperative content distribution: the intuition is that transmitting linear combinations, as opposed to original packets, makes distributed scheduling easier. Second, as discussed earlier in Section \ref{sec:wireless},  network coding works well with wireless broadcast: the intuition is that a coded packet broadcast over wireless is simultaneously useful to multiple receivers. Applications involving cooperative mobile devices combine these two features, cooperation and wireless transmission, and thus provide an ideal setting for harvesting the benefits of network coding.

Fortunately, network coding is  feasible in today's mobile phones, thanks to the increased computational and storage capabilities.
For example, the practicality of random network coding on iPhones has been discussed in \cite{RNC-iPHone}; the Tenor toolkit for network coding has been proposed for devices from servers to smartphones \cite{Tenor};  PictureViewer has demonstrated the feasibility of network coding on Nokia N95--8GB mobile phones \cite{PictureViewer}; NCUtils \cite{ncutils} is a C and Java library to implement network coding in various applications, and has recently been used as part of the Microcast  \cite{microcast-mobisys12} and MicroPlay \cite{microplay-mobigames12} frameworks on Android Nexus phones.

Due to both the need and opportunities for  network coding on cooperative mobile devices, there have been a number of recent research efforts that build such wireless systems primarily for media applications.  Next, we highlight some of these efforts, grouped according to their objective.

\subsubsection{Network coding for error recovery}

Downloading content, {\em e.g.}, through cellular or WiFi,  typically suffers from packet loss, due to noise and interference, on the downlink.  One possible solution is to have several devices, in close proximity of each other, cooperating for error recovery: they establish device-to-device connections and retransmit lost packets (or linear combinations of lost and original packets).   A rate-distortion optimized strategy  for cooperative video system repair using network coding in wireless peer-to-peer networks has been studied in \cite{RDO-NC-P2P}. Wireless video broadcasting with peer-to-peer error recovery has been proposed in \cite{Bopper} and an efficient scheduling approach to network coding for wireless local repair has been introduced in \cite{local_repair}. The authors in \cite{IPTV-VTC} further propose a cooperative IPTV system with pseudo-broadcast to improve reliability. In \cite{Hua_P2P,aalborg1}, base stations broadcast packets and smartphones establish device-to-device connections to help each other with error recovery.

\subsubsection{Network coding for delay}

Network coding can also help to reduce delay in wireless  settings. In \cite{GestureFlow}, the scenario of multiple users interacting through their smartphones using multi-touch gestures has been considered. A gesture broadcast protocol, called {\em GestureFlow}, is designed for concurrent gesture streams in multiple broadcast sessions using inter-session network coding. The goal of GestureFlow is to minimize the delay in recognizing a gesture in each session, which is critical for interactive applications.  An example application, called Music Score, has been developed to allow multiple users, communicating over a wide-area-network (WAN), to compose music together. It has been used for demonstration and evaluation of GestureFlow on iPads and iPods. However, the gesture streaming protocol itself is not specific to MusicScore and can work well with any application that requires intensive user interaction with multi-touch gestures.  The scenario differs from traditional multimedia streaming in that  gestures are very low rate, the communication scenario is all-to-all, and the gestures must be recovered with 100\% reliability and with very low delay and delay jitter.  The design of GestureFlow is based on the the following key features: adaptating the coding window, employing inter-session coding and using multiple paths.

In an ongoing work \cite{microplay-mobigames12}, the authors consider local multiplayer games on android phones. Similarly to GestureFlow,  the communication scenario is all-to-all and the metric to minimize is the delay in rendering all players' moves. Different from GestureFlow, however, all players are within proximity of each other. Current multiplayer games
 do not typically exploit the locality of the players. The authors in \cite{microplay-mobigames12}
develop a cooperative system using network coding, called {\em MicroPlay} that  exploits WiFi broadcast and network coding to reduce game latency and eliminate the need for input prediction by exploiting overheard packets.

\subsubsection{Network coding for throughput}

The authors in \cite{microcast-mobisys12,microcast-allerton11} consider  a group of smartphone users, within proximity of each other, who are interested in simulatenously viewing a video. The video can be stored either at a remote server or locally on one of the phones; an overview of this scenario is shown on Figure \ref{fig:smartphones}. A system, called {\em MicroCast}, has been designed using cooperation and network coding to maximize the common throughput to all users. Key ingredients of the MicroCast  design include the following. First, a scheduling algorithm, MicroDownload, decides what parts of the video each phone should download from the server, based on the download rates and the congestion in the local network. Second, a novel all-to-all local dissemination scheme, MicroNC-P2, is used for sharing content among group members.  MicroNC-P2  is specifically designed to exploit WiFi overhearing and network coding, based on a local high-rate WiFi broadcast mechanism, MicroBroadcast; it has been developed specifically for Android phones. MicroCast allows users to view the video simultaneously, while each user is able to stream video at a higher, and less varying, rate than if it would use only its single cellular connection or even state-of-the-art peer-to-peer schemes.
Although the current application used to demonstrate MicroCast is cooperative video streaming on mobile devices \cite{microcast-mobisys12}, the framework itself can be used for any application that requires high common throughput to a group of cooperative devices within proximity of each other.

\section{Conclusions}
\label{sec:conclusions}

We have reviewed in this paper the potentials and benefits of the network coding paradigm in multimedia applications. NC offers great advantages due to its ability to increase throughput as well as error resiliency, and to simplify the design of distributed architectures. These benefits perfectly match the growing demand for multimedia communications services over networks with sources, paths and clients diversity. We have discussed the adaptation of the NC principles to practical multimedia systems. An in-depth review of the recent NC research in peer-to-peer and wireless streaming systems has also been presented. NC offers important performance benefits in such popular settings where it increases the quality of service by effective exploitation of the network resources. In general, NC principles rely on the idea that in-network processing with actions close to bottlenecks or points of failures is key to the optimization of multimedia communication systems.

A few aspects have been omitted in our review of multimedia systems based on NC. In particular, this paper has not discussed the security issues posed by NC. NC security is an important topic on its own right and would deserve its own review paper. On the one hand, NC is more vulnerable than routing to byzantine modification attacks. Solutions to this vulnerability have been proposed with either error correction (following a typical communications and information theory approach) or novel cryptographic primitives (mainly from the cryptographic community). For example, a more detailed discussion can be found  in \cite{anh_le}. On the other hand, NC is more robust to eavesdropping attacks than a non-coded communication system. Quite a few research efforts have addressed these problems lately, although the proposed methods are not specific to multimedia systems.  In general, NC security methods are independent of the type of content shared through the NC system, {\em i.e.}, whether it is data or video, and thus outside the scope of this paper. The most relevant aspect to this review paper is the fact that, in video applications, pollution detection can be performed by the receivers at the video application layer \cite{Wang2010}. In addition, we have not considered the problem of distributed data storage where NC brings important benefits too. However, most of the research in this area is not focused on multimedia applications.

In summary, the most important research questions in the design of NC systems consider the effective control of delays and computational complexity in large scale distributed systems with multiple concurrent sessions. 
NC shares some similarity with channel coding and, in particular, rateless codes (which also use randomized coding strategies that are prone to distribution); it further complements distributed coding theory (which deals with distributed systems using a source coding approach). The development of both a unified theory and a solid practical framework in this area will certainly be of great benefit to the next generation of multimedia applications.


\end{document}